\documentclass[12pt,english]{article}
\usepackage{ae,aecompl}

\usepackage[T1]{fontenc}
\usepackage[utf8]{inputenc}
\usepackage{geometry}
\usepackage{babel}
\usepackage{amsmath}
\usepackage{amssymb}
\usepackage{esint}
\usepackage[unicode=true,pdfusetitle,
 bookmarks=true,bookmarksnumbered=false,bookmarksopen=false,
 breaklinks=false,pdfborder={0 0 1},backref=false,colorlinks=false]
 {hyperref}
\usepackage{breakurl}

\makeatletter
\numberwithin{equation}{section}

\usepackage{babel}
\usepackage{babel}
\usepackage{babel}
\usepackage{babel}
\usepackage{babel}
\usepackage{babel}
\usepackage{babel}
\usepackage{babel}
\usepackage{babel}
\newcommand{\ie}{{\it i.e.,} }

\makeatother

\begin{document}
\begin{flushright}
FIAN/TD/02-13\\

\par\end{flushright}

\vspace{1cm}

\begin{center}
\textbf{\Large{Off-Shell Scalar Supermultiplet in the Unfolded Dynamics
Approach}}{\Large{ \vspace{1cm}
}}
\par\end{center}{\Large \par}

\begin{center}
\textsc{N.G.Misuna$^{1}$ and M.A.Vasiliev$^{2}$} 
\par\end{center}

\begin{center}
\textsc{$^{1}$}\textit{Moscow Institute of Physics and Technology
(State University),}\\
 \textit{ Institutskii per. 9, 141700, Dolgoprudny, Moscow region,
Russia}\\

\par\end{center}

\begin{center}
\textit{$^{2}$I.E.Tamm Department of Theoretical Physics, P.N.Lebedev
Physical Institute,}\\
 \textit{ Leninsky prospect 53, 119991, Moscow, Russia}\\

\par\end{center}

\vspace{0.5cm}

\begin{abstract}
We show how manifestly supersymmetric action for Wess-Zumino model
can be constructed within the unfolded dynamics approach. The off-shell
unfolded system for ${\cal N}=1$, $D=4$ scalar supermultiplet is
found. The action is presented in the form of integral of a closed
4-form over any $(4,0)$ surface in superspace as well as a superspace
integral of an integral form or a chiral integral form. The proposed
method is argued to provide a most general tool for the analysis of
manifestly supersymmetric functionals. 
\end{abstract}
\newpage{}

\tableofcontents{}

\newpage{}

\section{Introduction}

Unfolded dynamics approach, originally developed for the description
of higher--spin field dynamics \cite{2 Vasiliev Annals Phys 190 (1989) 59},
implies rewriting field equations in the form of some generalized
covariant constancy conditions. In principle, any theory can be reformulated
in such a way (e.g., in \cite{3 Vasiliev Actions charges and off-shell fields in the unfolded dynamics approach}
this has been done for gravity and Yang-Mills theory). Unfolded formulation
of a dynamical system allows one to control its gauge symmetries.
The coordinate-free language of differential forms is particularly
convenient for theories of gravity. Moreover, so-called universal
unfolded equations \cite{4 Bekaert Cnockaert Iazeolla Vasiliev Nonlinear higher spin theories in various dimensions},
to which class belong all relevant examples, are insensitive to a
particular space-time where the fields live. The latter property allows
one in particular to derive different versions of the superspace formulations
directly from the unfolded formulation in the usual space-time.

A related remarkable feature of the unfolded dynamics approach is
that it provides a tool for the search for Lagrangians and conserved
currents in terms of certain $Q$-cohomology associated with the system
of unfolded equations \cite{3 Vasiliev Actions charges and off-shell fields in the unfolded dynamics approach}.
The aim of this paper is to illustrate this method by systematic derivation
of the manifestly supersymmetric superspace actions for the simplest
supersymmetric model, namely $4d$ Wess-Zumino model \cite{Wess Zumino 1974,18 Wess Bagger Supersymmetry and Supergravity}.
Our results provide an off-shell extension of the on-shell results
of \cite{Ponomarev Vasiliev}.

Naively, the proposed scheme may look obstructed by the fact that
superforms do not support integration over superspace. This is avoided
once, as we proceed in this paper, the action is either defined as
an integral of a superform over an even submanifold arbitrarily embedded
into the full superspace or as a Bernstein-Leites integral form \cite{Bernstein and Leites Integral Forms And The Stokes Formula On Supermanifolds}.
It should be noted that the Bernstein-Leites integration was applied
in the context of D-branes \cite{Howe Kappa-Symmetry} in which case
it was used that the integrands in the model in question had a specific
Gaussian form. In this paper we extend the class of integral forms
to those behaving as $\delta$-functions of the supervielbeins. The
respective superspace integrals turn out to be well-defined, providing
the manifestly supersymmetric formulation for supersymmetric theories.
Being useful in practical computation this extension of the class
of superspace Lagrangians may open new possibilities for the construction
of supersymmetric actions in superspace. In particular, for the Wess-Zumino
model considered in this paper integral forms having the form of $\delta$-functions
of supervielbeins provide explicit solutions for the equations that
determine invariant actions.

The results of this paper provide an example of the application of
the unfolded machinery to supersymmetric models, in which the superspace
constraints naturally arise via uplifting the unfolded system to the
superspace. We construct the most general unfolded Lagrangians of
the Wess-Zumino model in the form of a 4-superform, integral form
and a chiral integral form. They contain all possible supersymmetric
Lagrangians of the $4d$ Wess-Zumino model, \ie besides the standard
Wess-Zumino \cite{Wess Zumino 1974} and Salam-Strathdee \cite{Salam Strathdee}
Lagrangians (see also \cite{18 Wess Bagger Supersymmetry and Supergravity}),
they also contain higher-derivative Lagrangians. We expect that the
list of supersymmetric Lagrangians presented in this paper does not
go beyond the general supersymmetric Lagrangian $\mathcal{L}=\int d^{2}\theta d^{2}\bar{\theta}K+\left[\int d^{2}\theta W+h.c.\right]$
for the chiral superfield $\Phi$ ($\bar{D}_{\dot{\alpha}}\Phi=0$)
with arbitrary Lorentz-invariant superpotential $W(\Phi)$ and a (real)
Kahler potential $K$ which can depend on $\Phi$ and $\bar{\Phi}$
along with their (super)derivatives \cite{Weinberg SUSY}. Nevertheless
we believe that the obtained results can be useful for the practical
analysis of the higher-derivative component supersymmetric actions
since the manifestly supersymmetric superform unfolded action in our
approach directly reproduces the component action upon identification
of Minkowski space with an integration surface in the full superspace.
In the context of the Wess-Zumino model higher-derivative Lagrangians
were studied $\mathit{e.g.}$ in \cite{Antoniadis Dudas Ghilencea,Khoury Lehners Ovrut,Gallegos Senise da Silva,Gama Nascimento Petrov}.

In application to supersymmetric models, our method has much in common
with the group manifold approach \cite{D'Adda D'Auria Fre Regge,Castellani:1981um}
which treats actions as invariant functionals on hypersurface embedded
into the group manifold, as well as with the ``ectoplasm'' approach
of \cite{6 Gates Ectoplasm Has No Topology: The Prelude,7 Gates Grisaru Knutt-Wehlau Siegel Component Actions from Curved Superspace: Normal Coordinates and Ectoplasm,8 Gates Kuzenko Tartaglino-Mazzucchelli Chiral supergravity actions and superforms,Howe Brane Superemb}
in which supersymmetric component actions result from the $d$-closure
condition on the superLagrangian. The $Q$-cohomology used in the
unfolded dynamics is related to de Rham cohomology by virtue of unfolded
equations.

The nice feature of the $Q$-cohomology approach is that it reformulates
the problem in a coordinate-independent algebraic way, making the
procedure more systematic compared to the analysis of differential
operators as in the ectoplasm approach. As such, the $Q$-cohomology
approach is more general, being applicable to any (not necessarily
supersymmetric) model and/or geometry.

One of the key properties of the unfolded approach is that it makes
gauge and global symmetries manifest (the latter as residual symmetries
of the gauge symmetries of the field equations for the vacuum background
fields). In particular, local or global supersymmetry is one of such
symmetries depending on the model in question. It is this property
that makes the unfolded approach particularly useful for the study
of higher-spin gauge theories where it was originally developed. One
of the aims of this paper is to stress that it can be useful for the
analysis of usual (lower-spin) field theories as well. The feature
that this approach involves infinite towers of field variables sometimes
considered as a complication is, in fact, a simplification because
these towers of fields just form a basis of all off-shell or on-shell
nontrivial higher derivatives in the model. If one is not interested
in the analysis of certain higher derivatives it is possible to truncate
the unfolded equations appropriately. The benefit is that unfolded
formulation provides a well-defined basis of fields for the analysis
of higher-order and/or higher-derivative terms. The only condition
is the nilpotency of the operator $Q$, which guarantees both consistency
of the system and its symmetries. This condition provides a tool for
the search of manifestly supersymmetric formulation of the theory.
In this context, it should be noted that because the dynamical equations
of an unfolded system are classified in terms of the so-called $\sigma_{-}$-cohomology
\cite{5 Shaynkman and Vasiliev Theor. Math. Phys. 128 (2001) 1155},
this makes it possible to derive an off-shell unfolded system directly
from the on-shell one by finding such a completion of the unfolded
equations in which the corresponding sector of the $\sigma_{-}$-cohomology
vanishes. In principle, this provides a general approach for the search
of manifestly supersymmetric versions of the theories in question.

Though gauge symmetries do not play an essential role in the Wess-Zumino
model, we expect that the unfolded formalism may be useful for the
analysis of more complicated models like super Yang-Mills theory despite
the significant progress already achieved by different methods \cite{Berkovits,Bossard,Bossard2,Movshev,Chang 1,Chang 2}.
In this work we are mainly interested in the analysis of specificities
of the application of the unfolded technique to supersymmetric models
in superspace which is a modest step toward the future study of the
gauge supersymmetric theories.

The paper is organized as follows. In Section \ref{sec:2} unfolded
dynamics approach is overviewed and the cohomological method of computation
of Lagrangians for a graded system is proposed. In Section \ref{sec:3}
relevant aspects of superform integration are considered. In Section
\ref{sec:4} we recall the formulation of Minkowski superspace in
terms of flat superconnections. Section \ref{sec:5} contains an overview
of the results of \cite{Ponomarev Vasiliev} for the on-shell massless
scalar supermultiplet as well as its off-shell generalization. In
Section \ref{sec:6} we explore the operator $Q$ of the system in
question and compute the cohomology of its highest grade part. In
Section \ref{sec:Lagrangians} we derive and solve equations which
determine supersymmetric invariant functionals of the model and find
the particular solutions associated with action of the Wess-Zumino
model. In the end of Section \ref{sec:Lagrangians} it is explained
how conventional field-theoretic Lagrangians result from the found
unfolded Lagrangians and is argued that the latter describe all possible
supersymmetric Lagrangians in the model in question including the
standard expressions of \cite{18 Wess Bagger Supersymmetry and Supergravity}
along with their higher-derivative generalizations. Conventions and
notations are collected in \nameref{A}. The full system of equations
from Section \ref{sec:Lagrangians} is stored in \nameref{B}.

\section{Unfolded formulation \label{sec:2} }

\subsection{Unfolded equations\label{sub:2.1}}

Let $M^{d}$ be $d$-dimensional space-time manifold with local coordinates
$x^{\underline{n}}$, $\underline{n}=0,...,d-1$. Unfolding of equations
implies their reformulation in the form of generalized zero curvature
equations

\begin{equation}
R^{\Omega}(x):=dW^{\Omega}(x)+G^{\Omega}(W(x))=0,\label{eq:razv_ur-nie}
\end{equation}
where $d=dx^{\underline{m}}\tfrac{\partial}{\partial x^{\underline{m}}}$
is de Rham differential, $W^{\Omega}(x)$ are degree $p_{\Omega}$
differential forms and

\begin{equation}
G^{\Omega}(W^{\Upsilon}):=\sum_{n=1}^{\infty}f^{\Omega}{}_{\Upsilon_{1}...\Upsilon_{n}}W^{\Upsilon_{1}}...W^{\Upsilon_{n}}
\end{equation}
are degree $p_{\Omega}+1$ differential forms built from exterior
products of forms $W^{\Upsilon}(x)$ (wedge symbol is omitted in this
paper). Here $\Omega$ and $\Upsilon$ are indices carried by differential
forms.

The identity $d^{2}\equiv0$ implies the compatibility condition 
\begin{equation}
G^{\Upsilon}(W)\frac{\delta G^{\Omega}(W)}{\delta W^{\Upsilon}}\equiv0\,,\label{eq:razv_sovmestn}
\end{equation}
which has to be satisfied for all $W^{\Omega}$. It can be equivalently
rewritten as

\begin{equation}
Q^{2}=0,\qquad Q=G^{\Upsilon}(W)\frac{\delta}{\delta W^{\Upsilon}}\,.\label{eq:Q_sovm}
\end{equation}

Unfolded equations are called universal \cite{4 Bekaert Cnockaert Iazeolla Vasiliev Nonlinear higher spin theories in various dimensions,3 Vasiliev Actions charges and off-shell fields in the unfolded dynamics approach}
if compatibility condition (\ref{eq:razv_sovmestn}) holds independently
of the fact that any $p$-form with $p>d$ is zero in $d$-dimensional
space. In this case one can differentiate freely over $W^{\Omega}(x)$,
and equation (\ref{eq:razv_ur-nie}) is invariant under gauge transformation

\begin{equation}
\delta W^{\Omega}=d\varepsilon^{\Omega}-\varepsilon^{\Upsilon}\frac{\delta G^{\Omega}(W)}{\delta W^{\Upsilon}}\,,\label{eq:razv_kal_pr-e}
\end{equation}
where $(p_{\Omega}-1)$-form gauge parameter $\varepsilon^{\Omega}(x)$
is related to the $p_{\Omega}>0$ form $W^{\Omega}(x)$ (0-forms do
not give rise to gauge parameters).

For universal unfolded equations, condition (\ref{eq:razv_sovmestn})
holds independently of the choice of a space-time manifold. Full information
about local physical degrees of freedom of the unfolded system is
contained in 0-forms at any given point of space-time. Since these
data remain the same in any space, universal unfolded systems provide
an equivalent description in a larger (super)space simply via addition
of extra coordinates. Particular examples of this phenomenon have
been presented in \cite{10 Vasiliev Conformal Higher Spin Symmetries of 4d Massless Supermultiplets and Invariant Equations in Generalized (Super)Space,11 Engquist Sezgin Sundell Superspace Formulation of 4D Higher Spin Gauge Theory,13 Gelfond Vasiliev Higher Rank Conformal Fields in the $Sp(2M)$ Symmetric Generalized Space-Time,Ponomarev Vasiliev}.

The following terminology is used. The fields that can neither be
expressed via derivatives of some other fields nor gauged away are
called dynamical. The rest of the fields are referred to as auxiliary.
(Let us note that the decomposition of fields into dynamical and auxiliary
is not necessarily unambiguous). Differential conditions imposed by
unfolded equations on dynamical fields are called dynamical equations.
Other equations are either consequences of dynamical equations or
constraints which express auxiliary fields via derivatives of the
dynamical ones.

An example of unfolded equation can be constructed as follows. Let
$g$ be a Lie algebra with a basis $\left\{ T_{a}\right\} $. Consider
a $g$-valued 1-form $\Omega_{0}=\Omega_{0}^{a}T_{a}$. For $G=\Omega_{0}\Omega_{0}$,
equation (\ref{eq:razv_ur-nie}) reads as 
\begin{equation}
d\Omega_{0}+\Omega_{0}\Omega_{0}=0\,.\label{eq:dO+OO=00003D00003D0}
\end{equation}
The compatibility condition (\ref{eq:razv_sovmestn}) gives usual
Jacobi identity for the algebra $g$. Eq.~(\ref{eq:dO+OO=00003D00003D0})
means that the connection $\Omega_{0}$ is flat which is the standard
way to describe $g$-invariant vacuum. Eq.~(\ref{eq:razv_kal_pr-e})
gives usual gauge transformations of the connection $\Omega_{0}$

\begin{equation}
\delta\Omega_{0}=d\varepsilon_{0}\left(x\right)+\Omega_{0}\varepsilon_{0}\left(x\right)-\varepsilon_{0}\left(x\right)\Omega_{0},
\end{equation}
where $\varepsilon_{0}(x)$ is a 0-form valued in $g$. Given flat
connection $\Omega_{0}$ is invariant under the transformations with
parameters obeying

\begin{equation}
d\varepsilon_{0}\left(x\right)+\Omega_{0}\varepsilon_{0}\left(x\right)-\varepsilon_{0}\left(x\right)\Omega_{0}=0\,.\label{eq:glob_symm_eq}
\end{equation}
This equation is formally consistent by virtue of \eqref{eq:razv_sovmestn}.
Solutions of equations \eqref{eq:glob_symm_eq} describe the leftover
global symmetry $g$ of any solution of \eqref{eq:dO+OO=00003D00003D0}.

Let us linearize unfolded equations (\ref{eq:razv_ur-nie}) around
fixed connection $\Omega_{0}$ satisfying (\ref{eq:dO+OO=00003D00003D0}),
\[
W=\Omega_{0}+C,
\]
where $C$ are differential forms treated as small perturbations and
hence contributing linearly to the equations. Let $\left\{ C_{p}^{i}\right\} $
be a subset of forms of a fixed degree $p$, enumerated by index $i$.
In the linear approximation, the part of $G$ which is bilinear in
$\Omega_{0}$ and $C_{p}^{i}$ contributes, i.e. $G=\Omega_{0}^{a}\left(T_{a}\right)^{i}{}_{j}C_{p}^{j}$.
In this case, Eq.~(\ref{eq:razv_sovmestn}) implies that the matrices
$\left(T_{a}\right)^{i}{}_{j}$ form a representation of the algebra
$g$ in the space $V$ where $p$-forms $C_{p}^{i}$ are valued. Corresponding
equation (\ref{eq:razv_ur-nie}) is the covariant constancy condition
\begin{equation}
D_{\Omega_{0}}C_{p}^{i}=0\,,\label{eq: DC=00003D00003D0}
\end{equation}
where $D_{\Omega_{0}}\equiv d+\Omega_{0}$ is the covariant derivative
in the $g$-module $V$. $C_{p}^{i}$ transform properly under $g$
gauge transformations. Indeed, eq.~\eqref{eq:razv_kal_pr-e} gives
for \eqref{eq: DC=00003D00003D0}

\begin{equation}
\delta C_{p}^{i}=d\varepsilon_{p}^{i}-\varepsilon_{0}C_{p}^{i}+\Omega_{0}\varepsilon_{p}^{i}\,,
\end{equation}
where $\varepsilon_{p}^{i}$ are gauge parameters related to $C_{p}^{i}$
(for $p>0$) and $\varepsilon_{0}$ are global $g$-symmetry parameters
obeying \eqref{eq:glob_symm_eq}.

\subsection{$\sigma_{-}$-cohomology\label{sub:2.2}}

Classification of dynamical fields, gauge symmetries and dynamical
equations of the unfolded systems can be performed in terms of so-called
$\sigma_{-}$-cohomology \cite{5 Shaynkman and Vasiliev Theor. Math. Phys. 128 (2001) 1155,4 Bekaert Cnockaert Iazeolla Vasiliev Nonlinear higher spin theories in various dimensions,3 Vasiliev Actions charges and off-shell fields in the unfolded dynamics approach,Ponomarev Vasiliev}.
Let a linear unfolded system be of the form

\begin{equation}
\left(d+\underset{i}{{\displaystyle \sum}}\sigma_{i}\right)C(x)=0\,,\label{eq:(D+o)C=00003D00003D0}
\end{equation}
where $C(x)$ are some differential form fields and operators $\sigma_{i}$
act algebraically (i.e. do not differentiate $x^{\underline{n}}$).

In the $\sigma_{-}$- cohomology technics, the decomposition of the
fields into dynamical and auxiliary is controlled by the $\mathbb{Z}$-grading
$\mathcal{G}$ with respect to which auxiliary fields have higher
grade than dynamical ones. The grading operator $\mathcal{G}$ has
to be diagonalizable on the space of fields and to be bounded from
below. $d$ has grade zero. Usually, $\mathcal{G}$ counts a number
of tensor indices of the fields.

$\sigma_{-}$- cohomology technics applies if $\sigma_{i}$ contain
operators of negative grades. Then $\sigma_{-}$ is the operator of
the lowest grade and Eq.~(\ref{eq:(D+o)C=00003D00003D0}) takes the
form 
\begin{equation}
\left(d+\sigma_{-}+\Sigma\right)C(x)=0\label{eq:(D+o+S)C=00003D00003D0}
\end{equation}
with $\Sigma$ denoting all operators that act algebraically and have
$\mathcal{G}$-grade higher than $\sigma_{-}$. Since $\sigma_{-}$
has the lowest $\mathcal{G}$-grade, from compatibility condition
(\ref{eq:razv_sovmestn}) 
\begin{equation}
\left(d+\sigma_{-}+\Sigma\right)^{2}=0\label{eq:(D+o+S)^2=00003D00003D0}
\end{equation}
it follows that 
\begin{equation}
\left(\sigma_{-}\right)^{2}=0\,.
\end{equation}

Using that the gauge transformation (\ref{eq:razv_kal_pr-e}) for
equation (\ref{eq:(D+o+S)C=00003D00003D0}) is 
\begin{equation}
\delta C(x)=\left(d+\sigma_{-}+\Sigma\right)\varepsilon(x)\,,\label{eq:kal_pr-e_sigma}
\end{equation}
it can be shown \cite{5 Shaynkman and Vasiliev Theor. Math. Phys. 128 (2001) 1155,4 Bekaert Cnockaert Iazeolla Vasiliev Nonlinear higher spin theories in various dimensions,3 Vasiliev Actions charges and off-shell fields in the unfolded dynamics approach}
that, for $p$-forms $C_{p}$ from the space $V$, the cohomology
$H^{p-1}\left(\sigma_{-},V\right)$, $H^{p}\left(\sigma_{-},V\right)$
and $H^{p+1}\left(\sigma_{-},V\right)$ are, respectively, the spaces
of differential gauge symmetries, dynamical fields and dynamical equations.

The situation with several operators of negative grade is more complicated.
As shown in \cite{Ponomarev Vasiliev}, in this case usual $\sigma_{-}$-analysis
should be extended to the spectral sequence analysis of all such operators.
The full field-theoretical pattern of the system is determined by
the cohomology $H\left(\sigma_{-}^{'...'}|...|\sigma_{-}^{''}|\sigma_{-}^{'}|\sigma_{-}\right)$
where the operators $\sigma_{-}^{'...'}$ are arranged in the order
of increase of their $\mathcal{G}$-grade and $H\left(\sigma_{-}^{'}|\sigma_{-}\right)$
means the cohomology of $\sigma_{-}^{'}$ restricted to $H\left(\sigma_{-}\right)$.

\subsection{Unfolded actions and charges\label{sub:2.3}}

Invariants of a general unfolded system such as actions and conserved
charges are encoded by cohomology of the operator $Q$ (\ref{eq:Q_sovm})
\cite{3 Vasiliev Actions charges and off-shell fields in the unfolded dynamics approach}.

Suppose that system (\ref{eq:razv_ur-nie}) is off-shell, i.e. it
does not contain any dynamical equations, describing only a set of
constraints. In the language of $\sigma_{-}$-cohomology, this means
that unfolded equations for $\left(p-1\right)$-forms $W^{\Omega}$
have $H^{p}\left(\sigma_{-}\right)=0$. Following \cite{3 Vasiliev Actions charges and off-shell fields in the unfolded dynamics approach},
the action $S$ of this system is defined as an integral over a manifold
$M^{d}$ 
\begin{equation}
S=\underset{M^{d}}{\int}\mathcal{L}\label{eq:action}
\end{equation}
of some $d$-form $\mathcal{L}\left(W\right)$ which is a $Q$-closed
function of the fields $W^{\Omega}$ 
\begin{equation}
Q\mathcal{L}=0:\quad G^{\Upsilon}\left(W\right)\dfrac{\partial}{\partial W^{\Upsilon}}\mathcal{L}\left(W\right)=0\,.\label{eq:QL=00003D00003D0}
\end{equation}

Taking into account that $\delta\mathcal{L}=\left(\partial\mathcal{L}/\partial W^{\Omega}\right)\delta W^{\Omega}$
and using (\ref{eq:razv_kal_pr-e}), one easily obtains 
\begin{equation}
\delta\mathcal{L}=d\left(\varepsilon^{\Omega}\dfrac{\partial\mathcal{L}}{\partial W^{\Omega}}\right)\,.\label{eq:kal_pr-e_L-1}
\end{equation}
Assuming that $M^{d}$ has no boundary (or that fields decrease fast
enough at infinity), the action remains invariant under gauge transformations
(\ref{eq:razv_kal_pr-e}).

If the Lagrangian $\mathcal{L}$ is $Q$-exact, i.e. $\mathcal{L}=G^{\Omega}\tfrac{\partial\mathcal{F}}{\partial W^{\Omega}}$,
by virtue of (\ref{eq:razv_ur-nie}) 
\begin{equation}
\mathcal{L}=-dW^{\Omega}\dfrac{\partial\mathcal{F}}{\partial W^{\Omega}}=-d\mathcal{F}
\end{equation}
and hence $Q$-exact Lagrangians lead to trivial local actions. Thereby
nontrivial invariant actions of the off-shell system (\ref{eq:razv_ur-nie})
are in one-to-one correspondence with its $Q$-cohomology.

If system (\ref{eq:razv_ur-nie}) is on-shell (\ie contains some
dynamical equations) and a $p$-form $\mathcal{L}$ is a representative
of the nonzero $Q$-cohomology class, the same formula (\ref{eq:action})
describes a conserved charge as an integral over a $p$-cycle $\Sigma$
\begin{equation}
q=\underset{\Sigma}{\int}\mathcal{L}\,.
\end{equation}

Let $M^{d}$ be embedded into some ambient space, $M^{d}\subset{\widetilde{M}}^{\tilde{d}}$,
$\tilde{d}>d$. Extending \eqref{eq:razv_ur-nie} to ${\widetilde{M}}^{\tilde{d}}$,
by virtue of (\ref{eq:QL=00003D00003D0}), which is equivalent to
$d$-closure of $\mathcal{L}$, action \eqref{eq:action} is independent
of the local form of this embedding.

The case where the algebra of functions of fields from which a Lagrangian
is built admits a grading $G$ bounded from below, is of particular
interest. Let $Q$ and $\mathcal{L}$ admit decompositions into finite
sums of $G$-homogeneous parts 
\begin{equation}
Q={\displaystyle \sum_{i=0}^{n}}Q_{i}\,,\qquad\mathcal{L}={\displaystyle \sum_{i=0}^{k}}\mathcal{L}_{i}\,,
\end{equation}
where $G\left(Q_{i}\right)=G\left(\mathcal{L}_{i}\right)=i$. It can
be shown that the space of nontrivial $Q$-closed Lagrangians is isomorphic
to some subspace of $H(Q_{n})$, where $Q_{n}$ is the part of $Q$
of maximal $G$-grade.

Indeed, from $Q^{2}=0$ at different $G$-grades it follows that 
\begin{eqnarray}
 &  & \left(Q_{n}\right)^{2}=0,\label{eq:n^2=00003D0}\\
\nonumber \\
 &  & \left\{ Q_{n,}Q_{n-1}\right\} =0,\label{eq:{n,n-1}=00003D0}\\
 &  & ...\nonumber \\
 &  & \left(Q_{0}\right)^{2}=0.\label{eq:0^2=00003D0}
\end{eqnarray}
Equation $Q\mathcal{L}=0$ gives 
\begin{eqnarray}
 &  & Q_{n}\mathcal{L}_{k}=0,\label{eq:n_k=00003D0}\\
\nonumber \\
 &  & Q_{n-1}\mathcal{L}_{k}+Q_{n}\mathcal{L}_{k-1}=0,\label{eq:n-1_k=00003D0}\\
 &  & ...\nonumber \\
 &  & Q_{0}\mathcal{L}_{0}=0.\label{eq:0_0=00003D0}
\end{eqnarray}

Let the highest grade components be denoted as $\mathbb{Q}:=Q_{n}$
and $\mathbb{L}:=\mathcal{L}_{k}$. Since nontrivial Lagrangians are
represented by $Q$-cohomology, if $\mathbb{L}=\mathbb{Q}f$ it can
be removed by the redefinition 
\begin{equation}
\mathcal{L}'=\mathcal{L}-Qf
\end{equation}
so that $G(\mathbb{L}')<G(\mathbb{L})$, where $\mathbb{L}'$ is the
highest grade part of $\mathcal{L}'$. If $\mathbb{L}'=\mathbb{Q}g$,
the subtraction $\mathcal{L}''=\mathcal{L}'-Qg$ reduces the maximal
grade further. The process stops in a finite number of steps because
$G$-grading is bounded below. Eventually, either the Lagrangian vanishes
or its highest grade part $\mathbb{L}$ belongs to $H\left(\mathbb{Q}\right)$.
Thus, any nontrivial Lagrangian is represented by some $\mathbb{L}\in H\left(\mathbb{Q}\right)$.

This does not mean however that any $\mathbb{L}\in H\left(\mathbb{Q}\right)$
is associated with some nontrivial Lagrangian ${\mathcal{L}}\in H(Q)$.
Two related phenomena may happen.

One is that $\mathbb{L}$ cannot be supplemented with the terms of
the lowest degrees to form a $Q$-closed Lagrangian $\mathcal{L}$.
Indeed, $Q_{n-1}\mathbb{L}$ in Eq.~\eqref{eq:n-1_k=00003D0} is
$\mathbb{Q}$-closed by virtue of \eqref{eq:n^2=00003D0}, \eqref{eq:{n,n-1}=00003D0}
and \eqref{eq:n_k=00003D0}. If it is not $\mathbb{Q}$-exact however,
Eq.~\eqref{eq:n-1_k=00003D0} admits no solutions. In other words,
that $Q_{n-1}\mathbb{L}$ is in nontrivial $\mathbb{Q}$-cohomology
provides an obstruction for reconstruction of $\mathcal{L}$ in terms
of $\mathbb{L}$.

Another phenomenon is that if the same element of the $\mathbb{Q}$-cohomology,
that provides an obstruction for the extension to full $Q$-cohomology,
is interpreted as a highest grade part of some other Lagrangian with
$\mathbb{L}^{\prime}=Q_{n-1}\mathbb{L}$, then such a highest grade
part can be removed\textbf{ }by adding a $Q$-exact term -$Q\mathbb{L}$.

More generally, a similar phenomenon may occur at any step of the
analysis of Eqs.~\eqref{eq:n_k=00003D0}-\eqref{eq:0_0=00003D0}.
In particular, the corresponding $Q$-trivial highest grade components
have the form 
\begin{equation}
\mathbb{L}={\displaystyle \sum_{i=i_{0}}^{n}}Q_{i}\mathcal{F}_{k-i}=Q_{i_{0}}\mathcal{F}_{k-i_{0}}+Q_{i_{0}+1}\mathcal{F}_{k-i_{0}-1}+...+Q_{n-1}\mathcal{F}_{k-n+1}+\mathbb{Q}\mathcal{F}_{k-n},\label{eq:triv_L_1}
\end{equation}
with $\mathcal{F}_{i}$ obeying 
\begin{eqnarray}
 &  & Q_{i_{0}+1}\mathcal{F}_{k-i_{0}}+Q_{i_{0}+2}\mathcal{F}_{k-i_{0}-1}+...+\mathbb{Q}\mathcal{F}_{k-n+1}=0,\label{eq:triv_L_1.5}\\
 &  & ...\nonumber \\
 &  & Q_{n-1}\mathcal{F}_{k-i_{0}}+\mathbb{Q}\mathcal{F}_{k-i_{0}-1}=0,\\
\nonumber \\
 &  & \mathbb{Q}\mathcal{F}_{k-i_{0}}=0,\label{eq:triv_L_2}
\end{eqnarray}
where $i_{0}\in\left[0,n\right]$ is some fixed integer. If\textbf{
}$\mathcal{F}_{k-i_{0}}=\mathbb{Q}g$, it can be removed by the redefinition
$\mathcal{F}'_{i}=\mathcal{F}_{i}-Q_{n-k+i_{0}+i}g$ with $\mathcal{F}'_{i'}$
which, by virtue of \eqref{eq:n^2=00003D0}-\eqref{eq:0^2=00003D0},
obey the same system \eqref{eq:triv_L_1}-\eqref{eq:triv_L_2} with
$i'_{0}=i_{0}+1$ . As a result one is left either with a trivial
highest grade term $\mathbb{L}=\mathbb{Q}\widetilde{\mathcal{F}}_{k-n}$
(if all $\mathcal{F}_{i}$ in \eqref{eq:triv_L_1.5}-\eqref{eq:triv_L_2}
can be removed) or with $\widetilde{\mathcal{F}}_{k-i_{0}}$from \eqref{eq:triv_L_2}
that belongs to $H\left(\mathbb{Q}\right)$. In the latter case, though
being non-$\mathbb{Q}$-exact, $\mathbb{L}$ can be removed by adding
$Q$-exact terms $-\sum_{i}Q\mathcal{F}_{i}$.

Note that somewhat similar situation took place in \cite{Vasiliev Cubic Vertices},
where the deformation of Minkowski higher--spin vertices to $AdS$
space was studied. There nontrivial vertices belong to cohomology
of the nilpotent operator $Q=Q^{fl}+\lambda^{2}Q^{sub}$, where $-\lambda^{2}$
is the cosmological constant. The grading $G$ counts the number of
derivatives in vertices, $G\left(Q^{fl}\right)=1$, $G\left(Q^{sub}\right)=-1$.
The $AdS$ deformation (if exists) of a nontrivial vertex $F$ in
Minkowski space (where $\lambda=0$ and $Q=Q^{fl}$) may in principle
turn out to be trivial in $AdS$.

As a result, the space of nontrivial $Q$-closed Lagrangians is isomorphic
to subspace of $H\left(\mathbb{Q}\right)$, which is formed by some
highest grade terms $\mathbb{L}$ that cannot be represented in the
form \eqref{eq:triv_L_1} with $\mathcal{F}_{i}$ obeying \eqref{eq:triv_L_1.5}-\eqref{eq:triv_L_2}.

The construction of invariant functionals presented so far works nicely
for usual manifolds but is less obvious in the case of superspace
which is of most interest in this paper. Since, naively, the differential
superforms are not integrable over supermanifolds (see e.g., \cite{14 Witten Notes On Supermanifolds and Integration}),
we have to specify the notion of an unfolded action in superspace.

\section{Integration in superspace\label{sec:3}}

Most of differential geometry admits straightforward generalization
to supermanifolds. However extension of integration of differential
forms over a supermanifold is not quite straightforward. One way to
see this is to observe that superform transformation law does not
match Berezin integral. Indeed, consider a supermanifold $M^{p|q}$
with local coordinates $z^{M}=\left(x^{m},\theta^{\mu}\right)$. To
be coordinate-independent, the integration measure has to transform
according to Berezin formula 
\begin{equation}
\underset{M^{p|q}}{\int}f\left(x^{m},\theta^{\mu}\right)d^{p}xd^{q}\theta=\underset{M^{p|q}}{\int}f\left(y^{m},\xi^{\mu}\right)BerJd^{p}yd^{q}\xi\,,\label{eq:berez_integral}
\end{equation}
where 
\[
J=\left(\begin{array}{cc}
\dfrac{\partial x}{\partial y} & \dfrac{\partial x}{\partial\xi}\\
\dfrac{\partial\theta}{\partial y} & \dfrac{\partial\theta}{\partial\xi}
\end{array}\right)=\left(\begin{array}{cc}
A & B\\
C & D
\end{array}\right),\qquad BerJ=\dfrac{det(A-BD^{-1}C)}{detD}\,.
\]
Superforms resulting from naive extension to odd coordinates obviously
do not satisfy this condition and hence are not integrable.

However, even forms can still be integrated over even cycles in superspace.
Indeed, consider an even $n$-dimensional surface $\mathcal{S}^{n}$
on $M^{p|q}$: $z^{M}\left(t^{a}\right)=\left(x^{m}\left(t^{a}\right),\theta^{\mu}\left(t^{a}\right)\right)$,
$a=1,...,n$, parametrized by some even parameters $t^{a}$ (surface
coordinates). Let the integral of a $n$-superform $\omega=\omega_{M_{1}...M_{n}}dz^{M_{1}}...dz^{M_{n}}$
over $\mathcal{S}^{n}$ be defined as 
\begin{equation}
\underset{\mathcal{S}^{n}}{\int}\omega=\int\omega_{M_{1}...M_{n}}\left(\dfrac{\partial z^{M_{1}}}{\partial t^{a_{1}}}dt^{a_{1}}\right)...\left(\dfrac{\partial z^{M_{n}}}{\partial t^{a_{n}}}dt^{a_{n}}\right)=\int\omega'{}_{1...n}dt^{1}...dt^{n}\,.\label{eq:surf_integral}
\end{equation}
It is neither dependent on the choice of coordinates $t^{i}$ nor
of $z^{M}$. Obviously if $\omega$ is exact in superspace, the integration
over $\mathcal{S}^{n}$ amounts to that over its boundary $\partial\mathcal{S}^{n}$.
By Cartan formula ${\mathcal{L}}_{V}=\{d\,,i_{V}\}$ for the Lie derivative
of a vector field $V^{N}(z)$, the integral of a closed form $\omega$
($d\omega=0$) is independent of local variations of $\mathcal{S}^{n}$.
This allows us to define unfolded superfield action as the integral
of a closed $p$-superform $\mathcal{L}$ over an even $p$-dimensional
surface in superspace.

More generally, integration in superspace can be defined in terms
of integral forms, as was originally proposed in \cite{Bernstein and Leites Integral Forms And The Stokes Formula On Supermanifolds}
(see also \cite{14 Witten Notes On Supermanifolds and Integration}).
To this end an integral of superfunction $f\left(x,\theta\right)$
in superspace can be rewritten as 
\begin{equation}
\int f\left(x,\theta\right)d^{p}xd^{q}\theta=\int F\left(x,\theta,\varsigma,s\right)d^{p}xd^{q}\theta d^{p}\varsigma d^{q}s\,,\label{eq:int_F}
\end{equation}
where $\varsigma^{m}$, $m=1,...,p$ and $s^{\mu}$, $\mu=1,...,q$
are treated as additional anticommuting and commuting integration
variables, respectively. To contribute, $F\left(x,\theta,\varsigma,s\right)$
should have the form 
\begin{equation}
F\left(x,\theta,\varsigma,s\right)=f(x,\theta)\delta^{p}(\varsigma)\delta^{q}(s)\,.\label{eq:int_form}
\end{equation}
To make the link with differential forms, one formally substitutes
$dx^{m}$ and $d\theta^{\mu}$ for $\varsigma^{m}$ and $s^{\mu}$
in \eqref{eq:int_form}. Resulting objects are called integral forms.
Note that $\delta^{p}(dx)$ is just the usual volume form $dx^{1}\ldots dx^{p}$
while $\delta^{q}(d\theta)$ is the actual (even) $\delta$-function,
which is essentially non-polynomial. On the other hand, integration
of usual superforms polynomial in $d\theta$ does not make sense,
leading to divergent integral \eqref{eq:int_F} with $F\left(x,\theta,\varsigma,s\right)$
polynomial in $s^{\mu}$.

As mentioned in Introduction, our approach has much in common with
the ``ectoplasm'' method \cite{6 Gates Ectoplasm Has No Topology: The Prelude,7 Gates Grisaru Knutt-Wehlau Siegel Component Actions from Curved Superspace: Normal Coordinates and Ectoplasm,8 Gates Kuzenko Tartaglino-Mazzucchelli Chiral supergravity actions and superforms,Howe Brane Superemb}\textbf{
} of construction of manifestly supersymmetric actions represented
by integrals of superforms over space-time ``hypersurface'' in the
full superspace $M^{p|q}$ with coordinates $z^{M}=\left(x^{m},\theta^{\mu}\right)$.
Let a $p$-superform 
\begin{equation}
J=J_{M_{1}...M_{p}}dz^{M_{1}}...dz^{M_{p}}
\end{equation}
be closed 
\begin{equation}
dJ=0\quad\Rightarrow\quad\mathcal{D}_{[N}J_{M_{1...}M_{p})}-\dfrac{d}{2}T_{[NM|}{}^{P}J_{P|M_{2}..M_{p})}=0\,,\label{eq:dJ=00003D0}
\end{equation}
with the covariant derivative $\mathcal{D}_{M}$ and torsion tensor
$T_{MN}{}^{P}$. Then the integral over space-time hypersurface 
\begin{equation}
S=\underset{M^{p}}{\int}J_{m_{1}...m_{p}}dx^{m_{1}}...dx^{m_{p}}\label{eq:S_ectoplasm}
\end{equation}
is independent of coordinates in $M^{p|q}$ and, by virtue of \eqref{eq:dJ=00003D0},
of a particular choice of the integration hypersurface, provided that
the latter is even and $J$ falls down fast enough at spatial infinity.
It is invariant under the transformation 
\begin{equation}
\delta J_{M_{1}...M_{p}}=\partial_{[M_{1}}\lambda_{M_{2...}M_{p})}\,.
\end{equation}
The more general case of a curved superspace can be considered analogously
in terms of supervielbeins \cite{7 Gates Grisaru Knutt-Wehlau Siegel Component Actions from Curved Superspace: Normal Coordinates and Ectoplasm,8 Gates Kuzenko Tartaglino-Mazzucchelli Chiral supergravity actions and superforms}.
Note that in some applications of the ectoplasm approach the Bernstein-Leites
integration was also used \cite{Howe Kappa-Symmetry}. However, in
this case its applicability relied on the specific Gaussian form of
the integrands, allowing to carry out the integration over the odd
differentials. In this paper we will extend the application of the
Bernstein-Leites integration to distributions of the background supervielbeins
which provides a simple and efficient way for writing superinvariants.

Similarity of the ectoplasm and unfolded approaches is obvious. Indeed,
in the former $d$-closure of Lagrangian guarantees its manifest SUSY,
while in the latter the same is achieved via $Q$-closure condition.
As shown in Subsection \ref{sub:2.1}, $Q$ is an algebraic counterpart
of de Rham differential. Although for particular supersymmetric models
both methods lead to similar results, the unfolded approach is more
general being applicable to any (not necessarily supersymmetric) theory
and (generalized) space-time like $e.g.$, the $Sp(8)$-invariant
space-time considered in \cite{10 Vasiliev Conformal Higher Spin Symmetries of 4d Massless Supermultiplets and Invariant Equations in Generalized (Super)Space,13 Gelfond Vasiliev Higher Rank Conformal Fields in the $Sp(2M)$ Symmetric Generalized Space-Time}.

\section{Supersymmetric vacuum\label{sec:4}}

Following \cite{Ponomarev Vasiliev}, to obtain unfolded description
of the flat superspace we start with the $\mathcal{N}=1$ SUSY algebra
\begin{equation}
\left[M_{ab},M_{cd}\right]=-\left(\eta_{ac}M_{bd}+\eta_{bd}M_{ac}-\eta_{ad}M_{bc}-\eta_{bc}M_{ad}\right),\label{eq:SUSY_1}
\end{equation}

\begin{equation}
\left[P_{a},M_{bc}\right]=\eta_{ab}P_{c}-\eta_{ac}P_{b},
\end{equation}

\begin{equation}
\left\{ Q_{\alpha},\bar{Q}_{\dot{\beta}}\right\} =-2i\left(\sigma^{a}\right)_{\alpha\dot{\beta}}P_{a},
\end{equation}

\begin{equation}
\left[M_{ab},Q_{\alpha}\right]=\dfrac{1}{2}\left(\sigma_{ab}\right)_{\alpha}{}^{\beta}Q_{\beta},
\end{equation}

\begin{equation}
\left[M_{ab},\bar{Q}^{\dot{\alpha}}\right]=\dfrac{1}{2}\left(\bar{\sigma}_{ab}\right)^{\dot{\alpha}}{}_{\dot{\beta}}Q^{\dot{\beta}}.\label{eq:SUSY_2}
\end{equation}
(All other (anti)commutators are zero.)

Gauge fields of supergravity are 1-forms of vierbein $e^{a}=e_{\underline{m}}^{a}dx^{\underline{m}}$,
spin-connection $\omega^{a,b}=\omega_{\underline{m}}^{a,b}dx^{\underline{m}}$
and gravitino $\phi^{\alpha}=\phi_{\underline{m}}^{\alpha}dx^{\underline{m}}$
(see e.g. \cite{16 van Nieuwenhuizen Supergravity as a Yang-Mills theory}).
They are components of a 1-form connection $\Omega_{0}$ valued in
the SUSY algebra 
\begin{equation}
\Omega_{0}:=e^{a}P_{a}+\frac{1}{2}\omega^{a,b}M_{ab}+\phi^{\alpha}Q_{\alpha}+\bar{\phi}_{\dot{\alpha}}\bar{Q}^{\dot{\alpha}}\,.
\end{equation}

Supersymmetric flat background is represented by a connection $\Omega_{0}$
obeying \textbf{\eqref{eq:dO+OO=00003D00003D0}}, which leads to the
following component equations 
\begin{eqnarray}
 & D^{L}e^{a}+2i\phi^{\alpha}\bar{\phi}^{\dot{\alpha}}\left(\sigma^{a}\right)_{\alpha\dot{\alpha}}:=de^{a}+\omega^{a,b}e_{b}+2i\phi^{\alpha}\bar{\phi}^{\dot{\alpha}}\left(\sigma^{a}\right)_{\alpha\dot{\alpha}}=0,\label{eq:mink_tetr1}\\
\nonumber \\
 & D^{L}\omega^{a,b}:=d\omega^{a,b}+\omega^{a,c}\omega_{c}{}^{,b}=0,\\
\nonumber \\
 & D^{L}\phi^{\alpha}:=d\phi^{\alpha}+\dfrac{1}{4}\omega^{a,b}\phi^{\beta}\left(\sigma_{ab}\right)_{\beta}{}^{\alpha}=0,\\
\nonumber \\
 & D^{L}\bar{\phi}_{\dot{\alpha}}:=d\bar{\phi}_{\dot{\alpha}}+\dfrac{1}{4}\omega^{a,b}\bar{\phi}_{\dot{\beta}}\left(\bar{\sigma}_{ab}\right)^{\dot{\beta}}{}_{\dot{\alpha}}=0,\label{eq:mink_tetr2}
\end{eqnarray}
where $D^{L}\equiv d+\omega$ is the Lorentz covariant derivative.

As explained in Subsection \ref{sub:2.1}, to promote these unfolded
equations (which are obviously universal) to superspace it suffices
to add fermionic coordinates $x^{\underline{m}}\rightarrow z^{\underline{M}}=\left(x^{\underline{m}},\theta^{\underline{\mu}},\bar{\theta}^{\underline{\dot{\mu}}}\right)$
extending properly indices of the differential forms 
\begin{eqnarray*}
e_{\underline{m}}^{a}(x)dx^{\underline{m}}\rightarrow E_{\underline{M}}^{a}(z)dz^{\underline{M}}, &  & \omega_{\underline{m}}^{a,b}(x)dx^{\underline{m}}\rightarrow\Omega_{\underline{M}}^{a,b}(z)dz^{\underline{M}},\\
\\
\phi_{\underline{m}}^{\alpha}(x)dx^{\underline{m}}\rightarrow E_{\underline{M}}^{\alpha}(z)dz^{\underline{M}}, &  & \bar{\phi}_{\underline{m}}^{\dot{\alpha}}(x)dx^{\underline{m}}\rightarrow\bar{E}_{\underline{M}}^{\dot{\alpha}}(z)dz^{\underline{M}}.
\end{eqnarray*}
Flat superspace is described by the zero curvature equations 
\begin{eqnarray}
 & DE^{a}+2iE^{\alpha}\bar{E}^{\dot{\alpha}}\left(\sigma^{a}\right)_{\alpha\dot{\alpha}}:=dE^{a}+\Omega^{a,b}E_{b}+2iE^{\alpha}\bar{E}^{\dot{\alpha}}\left(\sigma^{a}\right)_{\alpha\dot{\alpha}}=0,\label{eq:SuSy_tetr1}\\
\nonumber \\
 & D\Omega^{a,b}:=d\Omega^{a,b}+\Omega^{a,c}\Omega_{c}{}^{,b}=0,\\
\nonumber \\
 & DE^{\alpha}:=dE^{\alpha}+\dfrac{1}{4}\Omega^{a,b}E^{\beta}\left(\sigma_{ab}\right)_{\beta}{}^{\alpha}=0,\\
\nonumber \\
 & D\bar{E}_{\dot{\alpha}}:=d\bar{E}_{\dot{\alpha}}+\dfrac{1}{4}\Omega^{a,b}\bar{E}_{\dot{\beta}}\left(\bar{\sigma}_{ab}\right)^{\dot{\beta}}{}_{\dot{\alpha}}=0,\label{eq:SuSy_tetr2}
\end{eqnarray}
where $D$ is the Lorentz covariant derivative in superspace.

Global SUSY transformations are described by those gauge transformations
\eqref{eq:glob_symm_eq} of system \eqref{eq:SuSy_tetr1}-\eqref{eq:SuSy_tetr2},
that leave invariant background connection 
\begin{eqnarray}
\delta E^{a} & = & d\varepsilon^{a}-\varepsilon^{a,b}E_{b}+\varepsilon_{b}\Omega^{a,b}-2i\left(\varepsilon^{\alpha}\bar{E}^{\dot{\alpha}}\left(\sigma^{a}\right)_{\alpha\dot{\alpha}}+\bar{\varepsilon}^{\dot{\alpha}}E^{\alpha}\left(\sigma^{a}\right)_{\alpha\dot{\alpha}}\right)=0,\label{eq:SuSy_kal_1}\\
\nonumber \\
\delta\Omega^{a,b} & = & d\varepsilon^{a,b}+\Omega^{a,c}\varepsilon_{c}\phantom{}^{b}-\varepsilon^{a,c}\Omega_{c}{}^{,b}=0,\\
\nonumber \\
\delta E^{\alpha} & = & d\varepsilon^{\alpha}+\frac{1}{4}\varepsilon^{\beta}\Omega^{a,b}\left(\sigma_{ab}\right)_{\beta}{}^{\alpha}-\frac{1}{4}\varepsilon^{a,b}E^{\beta}\left(\sigma_{ab}\right)_{\beta}{}^{\alpha}=0,\\
\nonumber \\
\delta\bar{E}_{\dot{\alpha}} & = & d\bar{\varepsilon}_{\dot{\alpha}}+\frac{1}{4}\Omega^{a,b}\bar{\varepsilon}_{\dot{\beta}}\left(\bar{\sigma}_{ab}\right)^{\dot{\beta}}{}_{\dot{\alpha}}-\frac{1}{4}\varepsilon^{a,b}\bar{E}_{\dot{\beta}}\left(\bar{\sigma}_{ab}\right)^{\dot{\beta}}{}_{\dot{\alpha}}=0\,.\label{eq:SuSy_kal_2}
\end{eqnarray}

Cartesian coordinate system is associated with the following solution
of \eqref{eq:SuSy_tetr1}-\eqref{eq:SuSy_tetr2} 
\[
E^{a}=dx^{\underline{m}}\delta_{\underline{m}}{}^{a}+d\theta^{\underline{\mu}}\left(i\bar{\theta}^{\dot{\underline{\mu}}}\left(\sigma^{a}\right)_{\underline{\mu}\underline{\dot{\mu}}}\right)+d\bar{\theta}_{\dot{\underline{\mu}}}\left(i\theta^{\underline{\mu}}\left(\sigma^{a}\right)_{\underline{\mu}\underline{\dot{\nu}}}\epsilon^{\underline{\dot{\mu}}\dot{\underline{\nu}}}\right),
\]

\begin{equation}
\Omega^{a,b}=0,\quad E^{\alpha}=d\theta^{\underline{\mu}}\delta_{\underline{\mu}}{}^{\alpha},\quad\bar{E}_{\dot{\alpha}}=d\bar{\theta}_{\dot{\underline{\mu}}}\delta^{\dot{\underline{\mu}}}{}_{\dot{\alpha}}\,.\label{eq:cartes_coord}
\end{equation}
In these coordinates, the explicit solution to system \eqref{eq:SuSy_kal_1}-\eqref{eq:SuSy_kal_2}
is 
\begin{eqnarray}
\varepsilon^{a} & = & \xi^{a}+\xi^{a}\phantom{}_{b}x^{b}+i\xi^{a}\phantom{}_{b}\theta^{\underline{\mu}}\bar{\theta}^{\dot{\underline{\mu}}}\left(\sigma^{b}\right)_{\underline{\mu}\underline{\dot{\mu}}}+2i\left(\xi^{\alpha}\bar{\theta}_{\dot{\underline{\mu}}}\left(\sigma^{a}\right)_{\alpha\dot{\alpha}}\epsilon^{\dot{\alpha}\dot{\underline{\mu}}}+\theta_{\underline{\mu}}\bar{\xi}^{\dot{\alpha}}\left(\sigma^{a}\right)_{\alpha\dot{\alpha}}\epsilon^{\underline{\mu}\alpha}\right),\\
\nonumber \\
\varepsilon^{a,b} & = & \xi^{a,b},\\
\nonumber \\
\varepsilon^{\alpha} & = & \frac{1}{4}\xi^{a,b}\theta_{\underline{\mu}}\epsilon^{\underline{\mu}\beta}\left(\sigma_{ab}\right)_{\beta}{}^{\alpha}+\xi^{\alpha},\label{eq:SuSy_kal_par-tr}\\
\nonumber \\
\bar{\varepsilon}_{\dot{\alpha}} & = & \frac{1}{4}\xi^{a,b}\bar{\theta}_{\underline{\dot{\mu}}}\delta^{\underline{\dot{\mu}}}\phantom{}_{\dot{\beta}}\left(\bar{\sigma}_{ab}\right)^{\dot{\beta}}{}_{\dot{\alpha}}+\bar{\xi}_{\dot{\alpha}}\label{eq:SuSy_kal_par-tr-1}
\end{eqnarray}
with $\xi^{a},\,\xi^{a,b}=-\xi^{b,a},\,\xi^{\alpha},\,\bar{\xi}_{\dot{\alpha}}$
being free constants, which are parameters of global symmetries.

Note that Lorentz covariant derivative in superspace can be rewritten
in the form 
\begin{equation}
D=E^{a}D_{a}+E^{\alpha}D_{\alpha}+\bar{E}_{\dot{\alpha}}D^{\dot{\alpha}}\,.\label{eq:D=00003DED+ED+ED}
\end{equation}
In Cartesian coordinates 
\begin{equation}
D_{a}=\partial_{a},\qquad D_{\alpha}=\partial_{\alpha}-i\left(\sigma^{b}\right)_{\alpha\dot{\alpha}}\bar{\theta}^{\dot{\alpha}}\partial_{b},\qquad\bar{D}_{\dot{\alpha}}=\bar{\partial}_{\dot{\alpha}}-i\theta^{\alpha}\left(\sigma^{b}\right)_{\alpha\dot{\alpha}}\partial_{b}\,.
\end{equation}

\section{Unfolded free massless scalar supermultiplet \label{sec:5}}

In this section we present an off-shell extension of the unfolded
equations of motion for $\mathcal{N}=1,$ $D=4$ free massless scalar
supermultiplet, obtained in \cite{Ponomarev Vasiliev}.

First we consider the problem in Minkowski space. It is described
by a solution of system (\ref{eq:mink_tetr1})-(\ref{eq:mink_tetr2})
with $\phi^{\alpha}=\bar{\phi}^{\dot{\alpha}}=0$, i.e. 
\begin{equation}
de^{a}+\omega^{a,b}e_{b}=0,\qquad d\omega^{a,b}+\omega^{a,c}\omega_{c}{}^{,b}=0\,.
\end{equation}

Massless scalar field in Minkowski space is described by the unfolded
equations \cite{2 Vasiliev Annals Phys 190 (1989) 59,5 Shaynkman and Vasiliev Theor. Math. Phys. 128 (2001) 1155}
\begin{equation}
D^{L}C^{a(k)}+e_{b}C^{a(k)b}=0\,,\label{eq:skal_pole}
\end{equation}
where the 0-forms $C^{a(k)}$ are symmetric traceless tensors of rank
$k$.

Similarly, unfolded equations 
\begin{equation}
D^{L}\chi_{\alpha}^{a(k)}+e_{b}\chi_{\alpha}^{a(k)b}=0\label{eq:spin_pole}
\end{equation}
for the complex 0-forms $\chi_{\alpha}^{a(k)}$ which are symmetric
traceless rank-$k$ spinor-tensors, obeying $\sigma$-transversality
condition 
\begin{equation}
\left(\bar{\sigma}_{b}\right)^{\dot{\alpha}\alpha}\chi_{\alpha}^{a(k-1)b}=0,
\end{equation}
describe massless spin-1/2 field in Minkowski space \cite{2 Vasiliev Annals Phys 190 (1989) 59,17 Vasiliev Higher Spin Superalgebras in any Dimension and their Representations}.

To unify systems (\ref{eq:skal_pole}) and (\ref{eq:spin_pole}) into
supermultiplet the terms with connections $\phi^{\alpha}$ and $\bar{\phi}^{\dot{\alpha}}$
which mix bosons and fermions have to be introduced. This gives \cite{Ponomarev Vasiliev}
\begin{eqnarray}
 &  & D^{L}C^{a(k)}+e_{b}C^{a(k)b}-\sqrt{2}\phi^{\alpha}\chi_{\alpha}^{a(k)}=0,\label{eq:C_ur-nie}\\
\nonumber \\
 &  & D^{L}\chi_{\alpha}^{a(k)}+e_{b}\chi_{\alpha}^{a(k)b}-\sqrt{2}i\bar{\phi}^{\dot{\alpha}}\left(\sigma_{b}\right)_{\alpha\dot{\alpha}}C^{a(k)b}=0\,.\label{eq:x_ur-nie}
\end{eqnarray}
Compatibility of the system is provided by flatness condition \eqref{eq:dO+OO=00003D00003D0}
and the identity $\left(\sigma_{b}\right)_{\beta\dot{\alpha}}\chi_{\alpha}^{a(k)b}=\left(\sigma_{b}\right)_{\alpha\dot{\alpha}}\chi_{\beta}^{a(k)b}$
which follows from $\sigma$-transversality and the fact that spinor
indices take two values.

Extension to superspace is trivially achieved via addition of fermionic
coordinates \textbf{$x^{\underline{m}}\rightarrow z^{\underline{M}}=\left(x^{\underline{m}},\theta^{\underline{\mu}},\bar{\theta}^{\underline{\dot{\mu}}}\right)$}
\begin{eqnarray}
 &  & DC^{a(k)}\left(z\right)+E_{b}C^{a(k)b}\left(z\right)-\sqrt{2}E^{\alpha}\chi_{\alpha}^{a(k)}\left(z\right)=0,\label{eq:SuSy_C_ur-nie}\\
\nonumber \\
 &  & D\chi_{\alpha}^{a(k)}\left(z\right)+E_{b}\chi_{\alpha}^{a(k)b}\left(z\right)-\sqrt{2}i\bar{E}^{\dot{\alpha}}\left(\sigma_{b}\right)_{\alpha\dot{\alpha}}C^{a(k)b}\left(z\right)=0\,.\label{eq:SuSy_x_ur-nie}
\end{eqnarray}
The resulting system is universal. As explained in Subsection \ref{sub:2.1},
this implies that all its symmetries are preserved. Hence, system
\eqref{eq:SuSy_C_ur-nie}, \eqref{eq:SuSy_x_ur-nie} is supersymmetric.

To check that these equations indeed describe free massless scalar
supermultiplet, one has to single out independent dynamical superfields
and dynamical equations with the help of $\sigma_{-}$-cohomology
technics. As shown in \cite{Ponomarev Vasiliev}, this gives the following
result. The only dynamical superfield is $C(z)$. All other fields
are auxiliary, being expressed via its derivatives. For instance,
$\chi_{\alpha}(z)=\frac{1}{\sqrt{2}}D_{\alpha}C(z)$. Independent
superfield equations are 
\begin{eqnarray}
 &  & \bar{D}_{\dot{\alpha}}C(z)=0,\label{eq:on-shell_1}\\
\nonumber \\
 &  & D^{\alpha}D_{\alpha}C(z)=0,\label{eq:on-shell_2}
\end{eqnarray}
which are standard equations of motion of a massless scalar supermultiplet
\cite{18 Wess Bagger Supersymmetry and Supergravity}.

To construct the action, we should find an off-shell modification
of system (\ref{eq:SuSy_C_ur-nie})-(\ref{eq:SuSy_x_ur-nie}) which
implies no dynamical equations. As a guiding example, first consider
the off-shell formulation of system \eqref{eq:skal_pole}, \eqref{eq:spin_pole}.
It results from relaxing the tracelessness condition for $C^{a(k)}$
and for $\chi_{\alpha}^{a(k)}$ as well as the $\sigma$-transversality
condition for the latter. Then Eqs.~(\ref{eq:skal_pole}), (\ref{eq:spin_pole})
just represent a set of constraints which express higher rank tensors
in terms of derivatives of the dynamical fields.

However, supersymmetric extension \eqref{eq:C_ur-nie}-\eqref{eq:x_ur-nie}
of the resulting off-shell system via introducing connections for
full SUSY algebra ceases to obey \eqref{eq:razv_sovmestn}, i.e. becomes
inconsistent. Indeed, the compatibility condition for Eq.~\eqref{eq:x_ur-nie}
requires 
\begin{equation}
\phi_{\alpha}\bar{\phi}_{\dot{\alpha}}\left(\bar{\sigma}_{b}\right)^{\dot{\alpha}\beta}\left(\chi^{a(k)b}\right)_{\beta}=0.
\end{equation}
In the on-shell case, it holds by virtue of $\sigma$-transversality,
which is relaxed in the off-shell case.

Inconsistency of the system means, in addition, that its gauge transformations
\eqref{eq:razv_kal_pr-e} do not obey SUSY algebra \eqref{eq:SUSY_1}-\eqref{eq:SUSY_2}
any more, i.e. the system lost SUSY. To restore both off-shell consistency
and SUSY, a set of auxiliary fields $F^{a(k)}$ should be introduced.
Supersymmetric off-shell system of equations acquires the form 
\begin{eqnarray}
 &  & D^{L}C^{a(k)}+e_{b}C^{a(k)b}-\sqrt{2}\phi^{\alpha}\chi_{\alpha}^{a(k)}=0,\label{eq: off-shell_mink_1}\\
\nonumber \\
 &  & D^{L}\chi_{\alpha}^{a(k)}+e_{b}\chi_{\alpha}^{a(k)b}-\sqrt{2}i\bar{\phi}^{\dot{\alpha}}\left(\sigma_{b}\right)_{\alpha\dot{\alpha}}C^{a(k)b}-\sqrt{2}\phi_{\alpha}F^{a(k)}=0,\\
\nonumber \\
 &  & D^{L}F^{a(k)}+e_{b}F^{a(k)b}-\sqrt{2}i\bar{\phi}_{\dot{\alpha}}\left(\bar{\sigma}_{b}\right)^{\dot{\alpha}\alpha}\chi_{\alpha}^{a(k)b}=0.\label{eq:off-shell_mink_2}
\end{eqnarray}
This system is consistent, obeying \eqref{eq:razv_sovmestn}. By virtue
of Eq.~\eqref{eq:off-shell_mink_2}, auxiliary fields $F^{a(k)}$
are higher derivatives of the ground auxiliary field $F$ familiar
for Wess-Zumino model. Not surprisingly, the fields $F^{a(k)}$ provide
closure of SUSY algebra of the off-shell system.

As explained in Subsection \ref{sub:2.1}, consistency of system \eqref{eq: off-shell_mink_1}-\eqref{eq:off-shell_mink_2}
implies its global SUSY invariance. Corresponding SUSY transformations
in Minkowski space are 
\begin{eqnarray}
\delta C^{a(k)} & = & \sqrt{2}\varepsilon^{\alpha}\chi_{\alpha}^{a(k)},\\
\nonumber \\
\delta\chi_{\alpha}^{a(k)} & = & \sqrt{2}i\bar{\varepsilon}^{\dot{\alpha}}\left(\sigma_{b}\right)_{\alpha\dot{\alpha}}C^{a(k)b}+\sqrt{2}\varepsilon_{\alpha}F^{a(k)},\\
\nonumber \\
\delta F^{a(k)} & = & \sqrt{2}i\bar{\varepsilon}_{\dot{\alpha}}\left(\bar{\sigma}_{b}\right)^{\dot{\alpha}\alpha}\chi_{\alpha}^{a(k)b}.
\end{eqnarray}

In Cartesian coordinates with $e^{a}\phantom{}_{\underline{m}}=\delta^{a}\phantom{}_{\underline{m}},\:\phi_{\alpha}=\bar{\phi}_{\dot{\alpha}}=0,\: D^{L}=d$
system \eqref{eq: off-shell_mink_1}-\eqref{eq:off-shell_mink_2}
implies $C_{a}=-\partial_{a}C$, $\left(\chi_{\alpha}\right)_{a}=-\partial_{a}\chi_{\alpha}$.
As a result, 
\begin{eqnarray}
\delta C & = & \sqrt{2}\xi^{\alpha}\chi_{\alpha},\\
\nonumber \\
\delta\chi_{\alpha} & = & -\sqrt{2}i\bar{\xi}^{\dot{\alpha}}\left(\sigma^{a}\right)_{\alpha\dot{\alpha}}\partial_{a}C+\sqrt{2}\xi_{\alpha}F,\\
\nonumber \\
\delta F & = & -\sqrt{2}i\bar{\xi}_{\dot{\alpha}}\left(\bar{\sigma}^{a}\right)^{\dot{\alpha}\alpha}\partial_{a}\chi_{\alpha},
\end{eqnarray}
where $\xi^{\alpha}$ and $\bar{\xi}^{\dot{\alpha}}$ are global SUSY
parameters. These are standard supertransformations of the chiral
supermultiplet \cite{18 Wess Bagger Supersymmetry and Supergravity}.

Extension of system \eqref{eq: off-shell_mink_1}-\eqref{eq:off-shell_mink_2}
to superspace is again achieved via extension of all functions to
superspace 
\begin{eqnarray}
 &  & DC^{a(k)}+E_{b}C^{a(k)b}-\sqrt{2}E^{\alpha}\chi_{\alpha}^{a(k)}=0,\label{eq: off-shell_1}\\
\nonumber \\
 &  & D\chi_{\alpha}^{a(k)}+E_{b}\chi_{\alpha}^{a(k)b}-\sqrt{2}i\bar{E}^{\dot{\alpha}}\left(\sigma_{b}\right)_{\alpha\dot{\alpha}}C^{a(k)b}-\sqrt{2}E_{\alpha}F^{a(k)}=0,\\
\nonumber \\
 &  & DF^{a(k)}+E_{b}F^{a(k)b}-\sqrt{2}i\bar{E}_{\dot{\alpha}}\left(\bar{\sigma}_{b}\right)^{\dot{\alpha}\alpha}\chi_{\alpha}^{a(k)b}=0.\label{eq:off-shell_2}
\end{eqnarray}
Resulting system imposes, however, differential equations with respect
to odd coordinates, i.e. strictly speaking it is not fully off-shell
in superspace. Indeed, using \eqref{eq:D=00003DED+ED+ED} it is easy
to obtain from \eqref{eq: off-shell_1}-\eqref{eq:off-shell_2} that

\begin{equation}
\bar{D}_{\dot{\alpha}}C^{a(k)}=0,\quad D_{\alpha}F^{a(k)}=0\,.\label{eq:DC=00003D0_DF=00003D0}
\end{equation}
These are chirality condition for the fields $C^{a(k)}$ and antichirality
condition for the fields $F^{a(k)}$. As is well known \cite{18 Wess Bagger Supersymmetry and Supergravity},
these conditions do not impose differential equations in Minkowski
space, where the system remains off-shell.

Since the fields of Wess-Zumino model are complex, system \eqref{eq: off-shell_1}-\eqref{eq:off-shell_2}
should be supplemented by the complex conjugated equations 
\begin{eqnarray}
 &  & D\bar{C}^{a(k)}+E_{b}\bar{C}^{a(k)b}+\sqrt{2}\bar{E}^{\dot{\alpha}}\bar{\chi}_{\dot{\alpha}}^{a(k)}=0,\label{eq:*off-shell_1}\\
\nonumber \\
 &  & D\bar{\chi}_{\dot{\alpha}}^{a(k)}+E_{b}\bar{\chi}_{\dot{\alpha}}^{a(k)b}+\sqrt{2}iE^{\alpha}\left(\sigma_{b}\right)_{\alpha\dot{\alpha}}\bar{C}^{a(k)b}-\sqrt{2}\bar{E}_{\dot{\alpha}}\bar{F}^{a(k)}=0,\\
\nonumber \\
 &  & D\bar{F}^{a(k)}+E_{b}\bar{F}^{a(k)b}-\sqrt{2}iE_{\alpha}\left(\bar{\sigma}_{b}\right)^{\dot{\alpha}\alpha}\bar{\chi}_{\dot{\alpha}}^{a(k)b}=0\,.\label{eq:*off-shell_2}
\end{eqnarray}
Their consequences

\begin{equation}
D_{\alpha}\bar{C}^{a(k)}=0,\quad\bar{D}_{\dot{\alpha}}\bar{F}^{a(k)}=0\label{eq:*DC=00003D0_*DF=00003D0}
\end{equation}
imply that $\bar{C}^{a(k)}$ are antichiral and $\bar{F}^{a(k)}$
are chiral.

\section{Operator $Q$\label{sec:6}}

\subsection{General properties}

According to the general scheme of \cite{3 Vasiliev Actions charges and off-shell fields in the unfolded dynamics approach}
recalled in Subsection \ref{sub:2.3}, Lagrangians of the unfolded
system are associated with its $Q$-cohomology. The full set of unfolded
equations of the system in question includes Eqs.~\eqref{eq:SuSy_tetr1}-\eqref{eq:SuSy_tetr2}
describing flat superspace background and Eqs.~\eqref{eq: off-shell_1}-\eqref{eq:off-shell_2},
\eqref{eq:*off-shell_1}-\eqref{eq:*off-shell_2} describing scalar
supermultiplet. The operator $Q$ of this system is 
\begin{equation}
Q=Q_{\Omega}+\hat{Q},
\end{equation}
where 
\begin{eqnarray}
Q_{\Omega} & = & \Omega^{a,b}E_{b}\dfrac{\partial}{\partial E^{a}}+\dfrac{1}{4}\Omega^{a,b}E^{\beta}\left(\sigma_{ab}\right)_{\beta}\phantom{}^{\alpha}\dfrac{\partial}{\partial E^{\alpha}}+\dfrac{1}{4}\Omega^{a,b}\bar{E}_{\dot{\beta}}\left(\bar{\sigma}_{ab}\right)^{\dot{\beta}}\phantom{}_{\dot{\alpha}}\dfrac{\partial}{\partial\bar{E}_{\dot{\alpha}}}+\Omega_{a,b}\hat{q}^{ba}+\nonumber \\
 &  & +\Omega^{a,c}\Omega_{c}\phantom{}^{,b}\dfrac{\partial}{\partial\Omega^{a,b}}\,,\label{eq:Q_omega}\\
\hat{Q} & = & 2iE^{\alpha}\left(\sigma^{a}\right)_{\alpha\dot{\alpha}}\bar{E}^{\dot{\alpha}}\dfrac{\partial}{\partial E^{a}}+E_{a}\hat{q}^{a}+\sqrt{2}E_{\alpha}\hat{q}^{\alpha}+\sqrt{2}\bar{E}_{\dot{\alpha}}\hat{\bar{q}}^{\dot{\alpha}}\,,\label{eq:Q0}
\end{eqnarray}
with 
\begin{eqnarray*}
\hat{q}^{b}\phantom{}_{c} & = & C^{a(k-1)b}\dfrac{\partial}{\partial C^{a(k-1)c}}+\bar{C}^{a(k-1)b}\dfrac{\partial}{\partial\bar{C}^{a(k-1)c}}+F^{a(k-1)b}\dfrac{\partial}{\partial F^{a(k-1)c}}+\bar{F}^{a(k-1)b}\dfrac{\partial}{\partial\bar{F}^{a(k-1)c}}+\\
 &  & +\chi_{\alpha}^{a(k-1)b}\dfrac{\partial}{\partial\chi_{\alpha}^{a(k-1)c}}-\dfrac{1}{4}\chi_{\beta}^{a(k)}\left(\sigma^{b}\phantom{}_{c}\right)^{\beta}\phantom{}_{\alpha}\dfrac{\partial}{\partial\chi_{\alpha}^{a(k)}}+\bar{\chi}_{\dot{\alpha}}^{a(k-1)b}\dfrac{\partial}{\partial\bar{\chi}_{\dot{\alpha}}^{a(k-1)c}}-\dfrac{1}{4}\bar{\chi}_{\dot{\beta}}^{a(k)}\left(\bar{\sigma}^{b}\phantom{}_{c}\right)^{\dot{\beta}}\phantom{}_{\dot{\alpha}}\dfrac{\partial}{\partial\bar{\chi}_{\dot{\alpha}}^{a(k)}}\,,\\
\\
\hat{q}^{b} & = & C^{a(k)b}\dfrac{\partial}{\partial C^{a(k)}}+\bar{C}^{a(k)b}\dfrac{\partial}{\partial\bar{C}^{a(k)}}+\chi_{\alpha}^{a(k)b}\dfrac{\partial}{\partial\chi_{\alpha}^{a(k)}}+\bar{\chi}_{\dot{\alpha}}^{a(k)b}\dfrac{\partial}{\partial\bar{\chi}_{\dot{\alpha}}^{a(k)}}+F^{a(k)b}\dfrac{\partial}{\partial F^{a(k)}}+\bar{F}^{a(k)b}\dfrac{\partial}{\partial\bar{F}^{a(k)}}\,,\\
\\
\hat{q}^{\alpha} & = & \left(\chi^{\alpha}\right)^{a(k)}\dfrac{\partial}{\partial C^{a(k)}}-F^{a(k)}\dfrac{\partial}{\partial\chi_{\alpha}^{a(k)}}-i\epsilon^{\alpha\beta}\left(\sigma_{b}\right)_{\beta\dot{\beta}}\bar{C}^{a(k)b}\dfrac{\partial}{\partial\bar{\chi}_{\dot{\beta}}^{a(k)}}-i\bar{\chi}_{\dot{\beta}}^{a(k)b}\left(\bar{\sigma}_{b}\right)^{\dot{\beta}\alpha}\dfrac{\partial}{\partial\bar{F}^{a(k)}}\,,\\
\\
\hat{\bar{q}}^{\dot{\alpha}} & = & -\left(\bar{\chi}^{\dot{\alpha}}\right)^{a(k)}\dfrac{\partial}{\partial\bar{C}^{a(k)}}+\bar{F}^{a(k)}\dfrac{\partial}{\partial\bar{\chi}_{\dot{\alpha}}^{a(k)}}-i\epsilon^{\dot{\alpha}\dot{\beta}}\left(\sigma_{b}\right)_{\beta\dot{\beta}}C^{a(k)b}\dfrac{\partial}{\partial\chi_{\beta}^{a(k)}}-i\left(\bar{\sigma}_{b}\right)^{\dot{\alpha}\beta}\chi_{\beta}^{a(k)b}\dfrac{\partial}{\partial F^{a(k)}}\,.
\end{eqnarray*}
$\hat{Q}$ can be represented in the form 
\begin{equation}
\hat{Q}=Q_{1}+Q_{2}^{+}+Q_{2}^{-}+Q_{3}\,,\label{eq:Q_ost}
\end{equation}
where the operators 
\begin{eqnarray}
 &  & Q_{1}=E_{a}\hat{q}^{a}\\
\nonumber \\
 &  & Q_{2}^{+}=\sqrt{2}E_{\alpha}\hat{q}^{\alpha},\quad Q_{2}^{-}=\sqrt{2}\bar{E}_{\dot{\alpha}}\hat{\bar{q}}^{\dot{\alpha}}\,,\\
\nonumber \\
 &  & Q_{3}=2iE^{\alpha}\left(\sigma^{a}\right)_{\alpha\dot{\alpha}}\bar{E}^{\dot{\alpha}}\dfrac{\partial}{\partial E^{a}}\,,
\end{eqnarray}
obey 
\begin{eqnarray}
 &  & \left(Q_{1}\right)^{2}=\left(Q_{2}^{+}\right)^{2}=\left(Q_{2}^{-}\right)^{2}=\left(Q_{3}\right)^{2}=0\,,\\
\nonumber \\
 &  & \left\{ Q_{1},Q_{3}\right\} =-\left\{ Q_{2}^{+},Q_{2}^{-}\right\} =2iE^{\alpha}\bar{E}^{\dot{\alpha}}\left(\sigma_{a}\right)_{\alpha\dot{\alpha}}\hat{q}^{a}\,,\label{eq:{Q+,Q-}}\\
\nonumber \\
 &  & \left\{ Q_{2}^{+},Q_{3}\right\} =\left\{ Q_{2}^{-},Q_{3}\right\} =\left\{ Q_{1},Q_{2}^{+}\right\} =\left\{ Q_{1},Q_{2}^{-}\right\} =0\,.
\end{eqnarray}
Eq.~\eqref{eq:{Q+,Q-}} implies

\begin{equation}
\{\hat{q}^{\alpha},\hat{\bar{q}}^{\dot{\alpha}}\}=-i\left(\bar{\sigma}_{a}\right)^{\dot{\alpha}\alpha}\hat{q}^{a}.\label{eq:{q+,q-}}
\end{equation}

\subsection{Highest grades\label{sub:high_grades}}

Now we are in a position to look for Lagrangians, representing cohomology
of the operator $Q$. These are built from background 1-forms $\Omega^{a,b}$,
$E_{a}$, $E_{\alpha}$, $\bar{E}_{\dot{\alpha}}$ and supermultiplet
0-form fields \textbf{$C^{a(k)},\:\bar{C}^{a(k)},\:\chi_{\alpha}^{a(k)},\:\bar{\chi}_{\dot{\alpha}}^{a(k)},\: F^{a(k)},\:\bar{F}^{a(k)}$}.

First of all, we observe that Lagrangian should be $\Omega$-independent
since the terms resulting from the action of $\Omega^{a,c}\Omega_{c}\phantom{}^{b}\dfrac{\partial}{\partial\Omega^{a,b}}$
in \eqref{eq:Q_omega} cannot be canceled against other terms. Indeed,
consider for instance a function $\Omega^{a,b}A_{ab}$, where $A_{ab}$
is built from 1-forms $E_{a},\: E_{\alpha},\:\bar{E}_{\dot{\alpha}}$
and 0-forms of supermultiplet fields. Then the part of $Q\Omega^{a,b}A_{ab}$
bilinear in $\Omega$ contains three terms of the form $\Omega^{a,c}\Omega_{c}\phantom{}^{b}A_{ab}$
which do not cancel. Similarly one proceeds with terms of higher orders
in $\Omega$. More precisely, though nonlinear $\Omega$-dependent
terms can be present, all of them can be removed by adding $Q$-exact
terms, thus representing the trivial class of $Q$-cohomology.

Clearly, for $\Omega$-independent Lagrangian, the condition 
\begin{equation}
Q_{\Omega}\mathcal{L}=0\label{eq:Q_omega_L=00003D00003D00003D00003D0}
\end{equation}
implies that $\mathcal{L}$ is Lorentz invariant, i.e. all indices
are contracted with Lorentz-invariant flat metric and $\sigma$-matrices.

As a result $Q$-cohomology amounts to cohomology of the $\Omega$-independent
part $\hat{Q}$ of $Q$. It is convenient to introduce the following
grading $G$ of the background 1-forms 
\begin{equation}
G\left(E_{\alpha}\right)=G\left(\bar{E}_{\dot{\alpha}}\right)=2,\quad G(E_{a})=1\,.
\end{equation}
Hence\textbf{ 
\begin{equation}
G\left(Q_{1}\right)=1,\quad G\left(Q_{2}^{+}\right)=G\left(Q_{2}^{-}\right)=2,\quad G\left(Q_{3}\right)=3,\,.
\end{equation}
} According to Subsection \ref{sub:2.3}, nontrivial Lagrangians can
be represented by cohomology $H\left(Q_{3}\right)$.

The computation considerably simplifies in spinor notations. The dictionary
between vector and spinor indices is provided by $\sigma$-matrices.
For example, for a Lorentz vector $A_{a}$ 
\[
A_{a}=\frac{1}{2}\left(\bar{\sigma}_{a}\right)^{\dot{\alpha}\alpha}A_{\alpha\dot{\alpha}},\qquad A_{\alpha\dot{\alpha}}=\left(\sigma_{a}\right)_{\alpha\dot{\alpha}}A^{a}.
\]
In spinor notations 
\begin{eqnarray}
 &  & Q_{1}=\dfrac{1}{2}E^{\alpha\dot{\alpha}}\hat{q}_{\alpha\dot{\alpha}},\\
\nonumber \\
 &  & Q_{3}=2i\bar{E}^{\dot{\alpha}}E^{\alpha}\dfrac{\partial}{\partial E^{\alpha\dot{\alpha}}}.\label{eq:Q3_spin}
\end{eqnarray}

An efficient tool to compute cohomology is provided by the homotopy
lemma (see e.g. \cite{Henneaux Teitelboimm Quantization of gauge systems}).
Consider $V={\textstyle \sum_{p=-\infty}^{\infty}}\bigoplus V^{p}$
where linear spaces $V^{p}$ are finite dimensional. Let $Q$ be a
grade one nilpotent operator 
\[
Q\left(V^{p}\right)\subset V^{p+1},\qquad Q^{2}=0,
\]
and $\widetilde{Q}$ be a grade -1 nilpotent operator 
\[
\widetilde{Q}\left(V^{p}\right)\subset V^{p-1},\qquad\widetilde{Q}^{2}=0.
\]
The homotopy lemma states, that if the homotopy operator 
\begin{equation}
\mathcal{H}:=\left\{ Q,\widetilde{Q}\right\} \label{eq:H}
\end{equation}
is diagonalizable in $V$, then cohomology $H\left(Q,V\right)\subset Ker\mathcal{H}$.
Indeed, let $v\in V$ be a $Q$-closed eigenvector of $\mathcal{H}$
\begin{equation}
\mathcal{H}v=\lambda v,\quad Qv=0
\end{equation}
with $\lambda\neq0$. Then $v$ is $Q$-exact because 
\begin{equation}
v=\lambda^{-1}\mathcal{H}v=Q\beta,\quad\beta=\lambda^{-1}\widetilde{Q}v.
\end{equation}
Hence, only $v$, that are eigenvectors of $\mathcal{H}$ with zero
eigenvalue, can belong to $H\left(Q,V\right)$.

To apply this technique let us introduce the operator 
\begin{equation}
\widetilde{Q}_{3}=\dfrac{1}{2i}E^{\alpha\dot{\alpha}}\dfrac{\partial^{2}}{\partial\bar{E}^{\dot{\alpha}}\partial E^{\alpha}},\quad\left(\widetilde{Q}_{3}\right)^{2}=0,\quad G\left(\widetilde{Q}_{3}\right)=-3\,.\label{eq:Q1_inv}
\end{equation}
The homotopy operator associated with \eqref{eq:Q3_spin} and \eqref{eq:Q1_inv}
is 
\begin{equation}
\mathcal{H}=E^{\alpha}\bar{E}^{\dot{\alpha}}\dfrac{\partial^{2}}{\partial\bar{E}^{\dot{\alpha}}\partial E^{\alpha}}+E^{\alpha\dot{\alpha}}\dfrac{\partial}{\partial E^{\alpha\dot{\alpha}}}+E^{\beta\dot{\alpha}}E^{\alpha}\dfrac{\partial^{2}}{\partial E^{\beta}\partial E^{\alpha\dot{\alpha}}}+E^{\alpha\dot{\beta}}\bar{E}^{\dot{\alpha}}\dfrac{\partial^{2}}{\partial\bar{E}^{\dot{\beta}}\partial E^{\alpha\dot{\alpha}}}.\label{eq:Hom_oper}
\end{equation}

As explained in Section \ref{sec:3}, in supersymmetric models one
can look for different types of unfolded superLagrangians, depending
on whether they are polynomials or distributions with respect to commuting
odd differentials $d\theta$ or, in terms of gauge fields, with respect
to gravitino 1-forms $E^{\alpha}$ and $\bar{E}^{\dot{\alpha}}$.
In the both cases $\mathcal{H}$ has zero $G$-grade, and is diagonalizable
so that the homotopy lemma applies.

It is convenient to characterize expressions in question by the number
of 1-forms $E^{\alpha\dot{\alpha}}$ they contain. Being anticommutative,
$E^{\alpha\dot{\alpha}}$ can appear only in the following combinations:
$E^{\alpha\dot{\alpha}}X_{\alpha\dot{\alpha}}$, $E^{\alpha}\phantom{}_{\dot{\alpha}}E^{\beta\dot{\alpha}}X_{\alpha\beta}+h.c.$,
$E^{\alpha}\phantom{}_{\dot{\alpha}}E^{\beta\dot{\alpha}}E_{\beta}\phantom{}^{\dot{\beta}}X_{\alpha\dot{\beta}},$
$E^{\alpha}\phantom{}_{\dot{\alpha}}E^{\beta\dot{\alpha}}E_{\beta\dot{\beta}}E_{\alpha}\phantom{}^{\dot{\beta}}X$,
where all $X$ are $E^{\alpha\dot{\alpha}}$-independent. Note that
for the last term in \eqref{eq:Hom_oper} the following relations
hold \textbf{ 
\begin{eqnarray}
 &  & E^{\alpha\dot{\beta}}\bar{E}^{\dot{\alpha}}\dfrac{\partial^{2}}{\partial\bar{E}^{\dot{\beta}}\partial E^{\alpha\dot{\alpha}}}\left(E^{\gamma}\phantom{}_{\dot{\gamma}}E^{\delta\dot{\gamma}}\bar{E}^{\dot{\delta}}\right)=E^{\gamma}\phantom{}_{\dot{\gamma}}E^{\delta\dot{\gamma}}\bar{E}^{\dot{\delta}},\label{eq:contr_ind}\\
\nonumber \\
 &  & E^{\alpha\dot{\beta}}\bar{E}^{\dot{\alpha}}\dfrac{\partial^{2}}{\partial\bar{E}^{\dot{\beta}}\partial E^{\alpha\dot{\alpha}}}\delta^{2}\left(\bar{E}_{\dot{\delta}}\right)E^{\gamma\dot{\gamma}}=-2\delta^{2}\left(\bar{E}_{\dot{\delta}}\right)E^{\gamma\dot{\gamma}},\label{eq:contr_ind_delta}
\end{eqnarray}
}using (A.6) and that $E^{\gamma}\phantom{}_{\dot{\gamma}}E^{\delta\dot{\gamma}}$
is symmetric in $\gamma$ and $\delta$ due to anticommutativity of
$E^{\alpha\dot{\alpha}}$. Derivative of $\delta$-function is defined
as usual via $\bar{E}_{\dot{\gamma}}\delta'_{\dot{\beta}}\left(\bar{E}_{\dot{\alpha}}\right)=\epsilon_{\dot{\gamma}\dot{\beta}}\delta^{2}\left(\bar{E}_{\dot{\alpha}}\right)$.
Analogous relations hold for the third term in \eqref{eq:Hom_oper}.

Now let us use the homotopy lemma to compute $Q_{3}$-cohomology in
the class of real superforms polynomial in $E^{\alpha}$ and $\bar{E}^{\dot{\alpha}}$.
Depending on the number of $E^{\alpha\dot{\alpha}}$, there are five
options: 
\begin{itemize}
\item $\Lambda_{0}=E^{\alpha}...E^{\alpha}\bar{E}^{\dot{\alpha}}...\bar{E}^{\dot{\alpha}}\lambda_{\alpha(m),\dot{\alpha}(n)}+h.c.,$
where $\lambda_{\alpha(m),\dot{\alpha}(n)}$ are 0-forms symmetric
over indices $\alpha$ and $\dot{\alpha}$. Here we have 
\begin{equation}
\mathcal{H}\Lambda_{0}=mn\Lambda_{0}+h.c.=0,
\end{equation}
which is true when $m=0$ or $n=0$, \textit{i.e.} 
\begin{equation}
\Lambda_{0}=E^{\alpha}...E^{\alpha}\lambda_{\alpha(m)}+h.c.\label{eq:H_1}
\end{equation}

\item $\Lambda_{1}=E^{\beta\dot{\beta}}E^{\alpha}...E^{\alpha}\bar{E}^{\dot{\alpha}}...\bar{E}^{\dot{\alpha}}\lambda_{\beta,\alpha(m),\dot{\beta},\dot{\alpha}(n)}+h.c.$

\begin{eqnarray}
\mathcal{H}\Lambda_{1} & = & mn\Lambda_{1}+\Lambda_{1}+mE^{\alpha\dot{\beta}}E^{\beta}E^{\alpha}...E^{\alpha_{m}}\bar{E}^{\dot{\alpha}}...\bar{E}^{\dot{\alpha}}\lambda_{\beta,\alpha(m),\dot{\beta},\dot{\alpha}(n)}+\nonumber \\
\nonumber \\
 &  & +nE^{\beta\dot{\alpha}}E^{\alpha}...E^{\alpha}\bar{E}^{\dot{\beta}}\bar{E}^{\dot{\alpha}}...\bar{E}^{\dot{\alpha}}\lambda_{\beta,\alpha(m),\dot{\beta},\dot{\alpha}(n)}+h.c.=0.
\end{eqnarray}
This equation has nontrivial solutions if $\beta$ is antisymmetrized
with some $\alpha_{1}$ in $\lambda_{\beta,\alpha(m),\dot{\beta},\dot{\alpha}(n)}$
(and similarly for dotted indices). This gives the following solutions

\end{itemize}
\begin{eqnarray}
\Lambda_{1}^{1} & = & E^{\alpha\dot{\alpha}}E_{\alpha}\bar{E}_{\dot{\alpha}}\lambda,\label{eq:H_2}\\
\nonumber \\
\Lambda_{1}^{2} & = & E^{\beta\dot{\alpha}}E_{\beta}E^{\alpha}...E^{\alpha}\lambda_{\alpha(m),\dot{\alpha}}+h.c.\label{eq:H_3}
\end{eqnarray}

\begin{itemize}
\item $\Lambda_{2}=E^{\beta}\phantom{}_{\dot{\beta}}E^{\gamma\dot{\beta}}E^{\alpha}...E^{\alpha}\bar{E}^{\dot{\alpha}}...\bar{E}^{\dot{\alpha}}\lambda_{\beta\gamma,\alpha(m),\dot{\alpha}(n)}+h.c.$

\begin{equation}
\mathcal{H}\Lambda_{2}=mn\Lambda_{2}+2\Lambda_{2}+2mE^{\alpha}\phantom{}_{\dot{\beta}}E^{\gamma\dot{\beta}}E^{\beta}E^{\alpha}...E^{\alpha}\bar{E}^{\dot{\alpha}}...\bar{E}^{\dot{\alpha}}\lambda_{\beta\gamma,\alpha(m),\dot{\alpha}(n)}+n\Lambda_{2}+h.c.=0.
\end{equation}
Again, this equation has nontrivial solutions only if $\lambda_{\beta,\gamma,\alpha(m),\dot{\alpha}(n)}$
is antisymmetrized over $\left(\beta,\alpha_{1}\right)$ and $\left(\gamma,\alpha_{2}\right)$,
\textit{i.e.}

\end{itemize}
\begin{equation}
\Lambda_{2}=E^{\beta}\phantom{}_{\dot{\alpha}}E^{\gamma\dot{\alpha}}E_{\beta}E_{\gamma}E^{\alpha}...E^{\alpha}\lambda_{\alpha(m)}+h.c.\label{eq:H_4}
\end{equation}

\begin{itemize}
\item $\Lambda_{3}=E^{\beta}\phantom{}_{\dot{\gamma}}E^{\gamma\dot{\gamma}}E_{\gamma}\phantom{}^{\dot{\beta}}E^{\alpha}...E^{\alpha}\bar{E}^{\dot{\alpha}}...\bar{E}^{\dot{\alpha}}\lambda_{\beta,\alpha(m),\dot{\beta},\dot{\alpha}(n)}+h.c.$

\begin{eqnarray}
\mathcal{H}\Lambda_{3} & = & mn\Lambda_{3}+3\Lambda_{3}+mE^{\alpha}\phantom{}_{\dot{\gamma}}E^{\gamma\dot{\gamma}}E_{\gamma}\phantom{}^{\dot{\beta}}E^{\beta}E^{\alpha}...E^{\alpha}\bar{E}^{\dot{\alpha}}...\bar{E}^{\dot{\alpha}}\lambda_{\beta,\alpha(m),\dot{\beta},\dot{\alpha}(n)}+m\Lambda_{3}+\nonumber \\
\nonumber \\
 &  & +nE^{\beta}\phantom{}_{\dot{\gamma}}E^{\gamma\dot{\gamma}}E_{\gamma}\phantom{}^{\dot{\alpha}}E^{\alpha}...E^{\alpha}\bar{E}^{\dot{\beta}}\bar{E}^{\dot{\alpha}}...\bar{E}^{\dot{\alpha}}\lambda_{\beta,\alpha(m),\dot{\beta},\dot{\alpha}(n)}+n\Lambda_{3}+h.c.=0.
\end{eqnarray}
This equation admits no nontrivial solutions.

\item $\Lambda_{4}=E^{\beta}\phantom{}_{\dot{\beta}}E^{\gamma\dot{\beta}}E_{\gamma\dot{\gamma}}E_{\beta}\phantom{}^{\dot{\gamma}}E^{\alpha}...E^{\alpha}\bar{E}^{\dot{\alpha}}...\bar{E}^{\dot{\alpha}}\lambda_{\alpha(m),\dot{\alpha}(n)}+h.c.$
From \eqref{eq:contr_ind} we find

\textbf{ 
\begin{eqnarray}
\mathcal{H}\Lambda_{4} & = & mn\Lambda_{4}+4\Lambda_{4}+2m\Lambda_{4}+2n\Lambda_{4}.
\end{eqnarray}
}

Hence, $\mathcal{H}\Lambda_{4}=0$ admits no nontrivial solutions.

\end{itemize}
Straightforward calculation shows that expressions \eqref{eq:H_1},
\eqref{eq:H_2}, \eqref{eq:H_3} and \eqref{eq:H_4} are $Q_{3}$-closed.
Thus cohomology of $Q_{3}$ in the class of superforms is contained
in 
\begin{eqnarray}
 &  & H_{1}=E^{\alpha}...E^{\alpha}\lambda_{\alpha(m)}+h.c.\label{eq:H1}\\
\nonumber \\
 &  & H_{2}=E^{\alpha\dot{\alpha}}E_{\alpha}\bar{E}_{\dot{\alpha}}\lambda,\\
\nonumber \\
 &  & H_{3}=E^{\beta\dot{\alpha}}E_{\beta}E^{\alpha}...E^{\alpha}\lambda_{\alpha(m),\dot{\alpha}}+h.c.\label{eq:H3}\\
\nonumber \\
 &  & H_{4}=E^{\beta}\phantom{}_{\dot{\alpha}}E^{\gamma\dot{\alpha}}E_{\beta}E_{\gamma}E^{\alpha}...E^{\alpha}\lambda_{\alpha(m)}+h.c.\label{eq:H4}
\end{eqnarray}
According to Section \ref{sec:3} nontrivial Lagrangians can be associated
with the 4-superforms from $H(Q_{3})$ which have the form

\begin{eqnarray}
 &  & \mathbb{L}_{8}=E^{\alpha}E^{\alpha}E^{\alpha}E^{\alpha}\ell_{\alpha(4)}+h.c.,\label{eq:L8}\\
\nonumber \\
 &  & \mathbb{L}_{7}=E^{\beta\dot{\alpha}}E_{\beta}E^{\alpha}E^{\alpha}\ell_{\alpha(2),\dot{\alpha}}+h.c.,\label{eq:L7}\\
\nonumber \\
 &  & \mathbb{L}_{6}=E^{\alpha}\phantom{}_{\dot{\alpha}}E^{\beta\dot{\alpha}}E_{\alpha}E_{\beta}\ell+h.c.,\label{eq:L0_polyn}
\end{eqnarray}
where $G\left(\mathbb{L}_{i}\right)=i$ and $\ell_{\alpha(4)}$, $\ell_{\alpha(2),\dot{\alpha}}$,
\textbf{$\ell$} are built from the supermultiplet fields. However,
it can be shown that \eqref{eq:L8} and \eqref{eq:L7} do not lead
to $Q$-closed expessions. Skipping details, here the phenomenon mentioned
in Subsection \ref{sub:2.3} occurs, namely $Q_{1}\mathbb{L}_{8}$
and $Q_{1}\mathbb{L}_{7}$ belong to nontrivial cohomology of $Q_{2}^{+}$,
that obstructs the reconstruction of the full Lagrangian.

So the only candidate for Lagrangians is \eqref{eq:L0_polyn}.\textbf{
}To single out trivial highest grade terms among \eqref{eq:L0_polyn}
consider the following 3-forms from $H(Q_{3})$ \eqref{eq:H1}-\eqref{eq:H4}
\begin{eqnarray}
\mathcal{F}_{6} & = & E^{\alpha}E^{\alpha}E^{\alpha}f_{\alpha(3)}+h.c.,\label{eq:F6}\\
\nonumber \\
\mathcal{F}_{5}^{1} & = & E^{\alpha\dot{\alpha}}E_{\alpha}\bar{E}_{\dot{\alpha}}f,\label{eq:F5_1}\\
\nonumber \\
\mathcal{F}_{5}^{2} & = & E^{\alpha\dot{\alpha}}E_{\alpha}E^{\beta}f_{\beta\dot{\alpha}}+h.c.\label{eq:F5_2}
\end{eqnarray}
where $G\left(\mathcal{F}_{i}\right)=i$ and $f$ are 0-forms built
from the supermultiplet fields. As shown in Subsection \ref{sub:2.3},
trivial Lagrangians are described by solutions of the system \eqref{eq:triv_L_1}-\eqref{eq:triv_L_2}
for \eqref{eq:L0_polyn} and some $\mathcal{F}$ \eqref{eq:F6}-\eqref{eq:F5_2}.

First, we observe that \eqref{eq:F6} cannot contribute since its
grade $G\left(\mathcal{F}_{6}\right)=6$ is the same as $\mathbb{L}_{6}$
\eqref{eq:L0_polyn}, so that the system \eqref{eq:triv_L_1}-\eqref{eq:triv_L_2}
admits no solutions since all $Q_{i}$ have positive $G$-grades.
For $\mathcal{F}_{5}^{1}$ \eqref{eq:F5_1} and $\mathcal{F}_{5}^{2}$
\eqref{eq:F5_2} with $G\left(\mathcal{F}_{5}^{1}\right)=G\left(\mathcal{F}_{5}^{2}\right)=5$
the system \eqref{eq:triv_L_1}-\eqref{eq:triv_L_2} takes the form
\begin{eqnarray}
 &  & Q_{3}\mathcal{F}_{5}^{i}=0,\label{eq:Q3_F=00003D0}\\
\nonumber \\
 &  & \left(Q_{2}^{+}+Q_{2}^{-}\right)\mathcal{F}_{5}^{i}+Q_{3}\mathcal{F}_{4}^{i}=0,\label{eq:Q2_F=00003D0}\\
\nonumber \\
 &  & Q_{1}\mathcal{F}_{5}^{i}+\left(Q_{2}^{+}+Q_{2}^{-}\right)\mathcal{F}_{4}^{i}+Q_{3}\mathcal{F}_{3}^{i}=E^{\alpha}\phantom{}_{\dot{\alpha}}E^{\beta\dot{\alpha}}E_{\alpha}E_{\beta}\ell+h.c.\label{eq:Q1_F=00003D0}
\end{eqnarray}
where $i=1,2$.

Eq.~\eqref{eq:Q3_F=00003D0} is satisfied since $\mathcal{F}_{5}^{1}$
and $\mathcal{F}_{5}^{2}$ are $Q_{3}$-closed. For $\mathcal{F}_{5}^{1}$,
Eq.~\eqref{eq:Q2_F=00003D0} admits a solution 
\begin{equation}
\mathcal{F}_{4}^{1}=\dfrac{i}{\sqrt{2}}E^{\alpha\dot{\alpha}}E^{\beta}\phantom{}_{\dot{\alpha}}\hat{q}_{\beta}f+h.c.\label{eq:F4_1}
\end{equation}
However \eqref{eq:Q1_F=00003D0} admits no nonzero solutions for $\mathcal{F}_{5}^{1}$
\eqref{eq:F5_1} and $\mathcal{F}_{4}^{1}$ \eqref{eq:F4_1} because
its l.h.s. contains among others the terms of the form 
\[
iE^{\alpha\dot{\alpha}}E^{\beta}\phantom{}_{\dot{\alpha}}\bar{E}^{\dot{\beta}}E_{\alpha}\hat{\bar{q}}_{\dot{\beta}}\hat{q}_{\beta}f+iE^{\alpha\dot{\alpha}}E_{\alpha}\phantom{}^{\dot{\beta}}E^{\beta}\bar{E}_{\dot{\alpha}}\hat{q}_{\beta}\hat{\bar{q}}_{\dot{\beta}}f
\]
that are not present on the r.h.s. of \eqref{eq:Q1_F=00003D0} and
cannot be compensated by any $Q_{3}\mathcal{F}_{3}^{1}$ due to \eqref{eq:{q+,q-}}
which in spinor notations reads as 
\begin{equation}
\{\hat{q}^{\alpha},\hat{\bar{q}}^{\dot{\alpha}}\}=-i\hat{q}^{\alpha\dot{\alpha}}.\label{eq:{q,q}_spin}
\end{equation}

For $\mathcal{F}_{5}^{2}$ \eqref{eq:F5_2}, Eq.~\eqref{eq:Q2_F=00003D0}
gives 
\begin{eqnarray}
 &  & \hat{q}_{\alpha}f_{\beta\dot{\alpha}}=0,\qquad\hat{\bar{q}}_{\dot{\alpha}}\bar{f}_{\dot{\beta}\alpha}=0,\label{eq:f_kir}\\
\nonumber \\
 &  & \mathcal{F}_{4}^{2}=i\sqrt{2}E^{\alpha}\phantom{}_{\dot{\alpha}}E^{\beta\dot{\alpha}}E_{\alpha}\hat{\bar{q}}^{\dot{\beta}}f_{\beta\dot{\beta}}+h.c.
\end{eqnarray}
Then \eqref{eq:Q1_F=00003D0} is solved by 
\begin{eqnarray}
 &  & \mathcal{F}_{3}^{2}=-\frac{2}{3}E^{\alpha}\phantom{}_{\dot{\beta}}E^{\beta\dot{\beta}}E_{\beta}\phantom{}^{\dot{\alpha}}\hat{\bar{q}}_{\dot{\gamma}}\hat{\bar{q}}^{\dot{\gamma}}f_{\alpha\dot{\alpha}}+h.c.,\\
\nonumber \\
 &  & \ell=-\frac{1}{2}\hat{q}^{\alpha\dot{\alpha}}f_{\alpha\dot{\alpha}}.\label{eq:l_triv}
\end{eqnarray}

The conclusion is that Lagrangians generated by $\mathbb{L}$ \eqref{eq:L0_polyn}
are trivial if $\ell$ has the form \eqref{eq:l_triv} (analogously
for $\bar{\ell}$) with $f_{\alpha\dot{\alpha}}$ obeying \eqref{eq:f_kir}.

\section{Lagrangians\label{sec:Lagrangians}}

\subsection{Four-form Lagrangian\label{sub:4-form}}

We look for a Lagrangian as a $Q$-closed 4-superform 
\begin{eqnarray}
\mathcal{L} & = & E_{a}E_{b}\left\{ \left(\bar{\sigma}^{ab}\right)^{\dot{\alpha}\dot{\beta}}\bar{E}_{\dot{\alpha}}\bar{E}_{\dot{\beta}}\ell_{6}+\left(\sigma^{ab}\right)^{\alpha\beta}E_{\alpha}E_{\beta}\bar{\ell}_{6}\right\} +\nonumber \\
 &  & +\epsilon^{abcd}E_{a}E_{b}E_{c}\left\{ \bar{E}_{\dot{\alpha}}\left(\bar{\sigma}_{d}\right)^{\dot{\alpha}\alpha}\ell_{5}\phantom{}_{\alpha}+E^{\alpha}\left(\sigma_{d}\right)_{\alpha\dot{\alpha}}\bar{\ell}_{5}^{\dot{\alpha}}\right\} +E_{a}E_{b}E_{c}E_{d}\epsilon^{abcd}\ell_{4},\label{eq:ansatz_L}
\end{eqnarray}
where 0-forms $\ell_{i}$ are built Lorentz-covariantly from the supermultiplet
fields \textbf{$C^{a(k)},$} $\bar{C}^{a(k)},$ $\chi_{\alpha}^{a(k)},$
$\bar{\chi}_{\dot{\alpha}}^{a(k)},$ $F^{a(k)},$ $\bar{F}^{a(k)}$.
That this is the most general Lorentz-invariant 4-superform Ansatz
follows from the results of Section \ref{sub:high_grades}. (For instance,
the term $\epsilon^{abcd}E_{a}E_{b}E^{\alpha}\left(\sigma_{c}\right)_{\alpha\dot{\alpha}}\bar{E}^{\dot{\alpha}}\ell_{d}$,
that has the same $G$-grade as \eqref{eq:L0_polyn}, is not added
as not representing nonzero $Q_{3}$-cohomology.)

The equation $\hat{Q}\mathcal{L}=0$ can now be analyzed in different
grade sectors, starting from the highest one. The full set of equations
is presented in \nameref{B}.

There are two complex conjugated equations in the highest grade $G=9$
\begin{eqnarray}
2iE^{\gamma}\left(\sigma^{c}\right)_{\gamma\dot{\gamma}}\bar{E}^{\dot{\gamma}}\dfrac{\partial}{\partial E^{c}}E_{a}E_{b}\left(\bar{\sigma}^{ab}\right)^{\dot{\alpha}\dot{\beta}}\bar{E}_{\dot{\alpha}}\bar{E}_{\dot{\beta}}\ell_{6} & = & 0,\label{eq:G=00003D9}\\
\nonumber \\
2iE^{\gamma}\left(\sigma^{c}\right)_{\gamma\dot{\gamma}}\bar{E}^{\dot{\gamma}}\dfrac{\partial}{\partial E^{c}}E_{a}E_{b}\left(\sigma^{ab}\right)^{\alpha\beta}E_{\alpha}E_{\beta}\bar{\ell}_{6} & = & 0\,.
\end{eqnarray}
These hold true for any $\ell_{6}$ and $\bar{\ell}_{6}$, because
the corresponding terms belong to $H\left(Q_{3}\right)$. Indeed,
e.g. Eq.~\eqref{eq:G=00003D9} is proportional to $\epsilon^{abcd}\left(\bar{\sigma}_{ab}\right)^{\dot{\alpha}\dot{\beta}}\left(\bar{\sigma}_{c}\right)^{\dot{\gamma}\alpha}\bar{E}_{\dot{\alpha}}\bar{E}_{\dot{\beta}}\bar{E}_{\dot{\gamma}}$.
From relation (A.9) it follows that to be symmetric over three dotted
spinor indices $\epsilon^{abcd}\left(\bar{\sigma}_{ab}\right)^{\dot{\alpha}\dot{\beta}}\left(\bar{\sigma}_{c}\right)^{\dot{\gamma}\alpha}$
must be antisymmetric with respect to the three undotted indices,
which is zero because the latter take just two values.

In the grade $G=8$ we obtain four equations. From the first two 
\begin{eqnarray}
\bar{E}_{\dot{\gamma}}\hat{\bar{q}}^{\dot{\gamma}}E_{a}E_{b}\left(\bar{\sigma}^{ab}\right)^{\dot{\alpha}\dot{\beta}}\bar{E}_{\dot{\alpha}}\bar{E}_{\dot{\beta}}\ell_{6} & = & 0,\\
\nonumber \\
E_{\gamma}\hat{q}^{\gamma}E_{a}E_{b}\left(\sigma^{ab}\right)^{\alpha\beta}E_{\alpha}E_{\beta}\bar{\ell}_{6} & = & 0,
\end{eqnarray}
we find that $\ell_{6}$ and $\bar{\ell}_{6}$ have to obey $\hat{\bar{q}}_{\dot{\alpha}}\ell_{6}=0$
and $\hat{q}_{\alpha}\bar{\ell}_{6}=0$, respectively. Using (A.9),
one easily finds that the last two equations 
\begin{eqnarray}
 &  & 2iE^{\gamma}\left(\sigma^{e}\right)_{\gamma\dot{\gamma}}\bar{E}^{\dot{\gamma}}\dfrac{\partial}{\partial E^{e}}\epsilon^{abcd}E_{a}E_{b}E_{c}\bar{E}_{\dot{\alpha}}\left(\bar{\sigma}_{d}\right)^{\dot{\alpha}\alpha}\ell_{5}\phantom{}_{\alpha}+\sqrt{2}E_{\gamma}\hat{q}^{\gamma}E_{a}E_{b}\left(\bar{\sigma}^{ab}\right)^{\dot{\alpha}\dot{\beta}}\bar{E}_{\dot{\alpha}}\bar{E}_{\dot{\beta}}\ell_{6}=0,\\
\nonumber \\
 &  & 2iE^{\gamma}\left(\sigma^{e}\right)_{\gamma\dot{\gamma}}\bar{E}^{\dot{\gamma}}\dfrac{\partial}{\partial E^{e}}\epsilon^{abcd}E_{a}E_{b}E_{c}E^{\alpha}\left(\sigma_{d}\right)_{\alpha\dot{\alpha}}\bar{\ell}_{5}^{\dot{\alpha}}+\sqrt{2}\bar{E}_{\dot{\gamma}}\hat{\bar{q}}^{\dot{\gamma}}E_{a}E_{b}\left(\sigma^{ab}\right)^{\alpha\beta}E_{\alpha}E_{\beta}\bar{\ell}_{6}=0
\end{eqnarray}
are solved by $\ell_{5}\phantom{}_{\alpha}=\dfrac{\sqrt{2}}{6}\hat{q}_{\alpha}\ell_{6}$
and $\bar{\ell}_{5}\phantom{}_{\dot{\alpha}}=\dfrac{\sqrt{2}}{6}\hat{\bar{q}}_{\dot{\alpha}}\bar{\ell}_{6}$.

Continuation of this analysis gives the following $Q$-closed superfield
Lagrangian 
\begin{eqnarray}
\mathcal{L} & = & E_{a}E_{b}\left(\bar{\sigma}^{ab}\right)^{\dot{\alpha}\dot{\beta}}\bar{E}_{\dot{\alpha}}\bar{E}_{\dot{\beta}}W+E_{a}E_{b}\left(\sigma^{ab}\right)^{\alpha\beta}E_{\alpha}E_{\beta}\bar{W}+\nonumber \\
\nonumber \\
 &  & +\dfrac{\sqrt{2}}{6}\epsilon^{abcd}E_{a}E_{b}E_{c}\bar{E}_{\dot{\alpha}}\left(\bar{\sigma}_{d}\right)^{\dot{\alpha}\alpha}\hat{q}_{\alpha}W+\dfrac{\sqrt{2}}{6}\epsilon^{abcd}E_{a}E_{b}E_{c}E^{\alpha}\left(\sigma_{d}\right)_{\alpha\dot{\alpha}}\hat{\bar{q}}^{\dot{\alpha}}\bar{W}+\nonumber \\
\nonumber \\
 &  & +E_{a}E_{b}E_{c}E_{d}\epsilon^{abcd}\left(\dfrac{i\sqrt{2}}{16}\hat{\bar{q}}_{\dot{\alpha}}\hat{\bar{q}}^{\dot{\alpha}}\bar{W}-\dfrac{i\sqrt{2}}{16}\hat{q}^{\alpha}\hat{q}_{\alpha}W\right)\label{eq:gen_lagr}
\end{eqnarray}
where the 0-forms $W$ and $\bar{W}$ are arbitrary functions of $C^{a(k)}$,
$\bar{F}^{a(k)}$ and \textbf{$\bar{C}^{a(k)}$ }, $F^{a(k)}$, respectively,
so that $\hat{\bar{q}}^{\dot{\alpha}}W=0$ and $\hat{q}^{\alpha}\bar{W}=0$.
By virtue of \eqref{eq:DC=00003D0_DF=00003D0}, \eqref{eq:*DC=00003D0_*DF=00003D0}
this means that $W$ is chiral and $\bar{W}$ is antichiral. Note
that, as follows from \eqref{eq:l_triv}, $W$ of the form $W=\hat{q}_{a}f^{a}$
lead to trivial Lagrangians.

By construction, Lagrangian \eqref{eq:gen_lagr} is manifestly supersymmetric
and the corresponding action is independent of the local variation
of the integration surface. A particular solution, that reproduces
free Wess-Zumino action \cite{18 Wess Bagger Supersymmetry and Supergravity},
results from \eqref{eq:gen_lagr} with $W=i2\sqrt{2}C\bar{F}$, $\bar{W}=-i2\sqrt{2}\bar{C}F$
\begin{eqnarray}
\mathcal{L^{WZ}} & = & i2\sqrt{2}E_{a}E_{b}\left(\bar{\sigma}^{ab}\right)^{\dot{\alpha}\dot{\beta}}\bar{E}_{\dot{\alpha}}\bar{E}_{\dot{\beta}}C\bar{F}-i2\sqrt{2}E_{a}E_{b}\left(\sigma^{ab}\right)^{\alpha\beta}E_{\alpha}E_{\beta}\bar{C}F+\nonumber \\
\nonumber \\
 &  & +\dfrac{2i}{3}\epsilon^{abcd}E_{a}E_{b}E_{c}\bar{E}_{\dot{\alpha}}\left(\bar{\sigma}_{d}\right)^{\dot{\alpha}\alpha}\left[\chi_{\alpha}\bar{F}-iC\left(\bar{\sigma}^{e}\right)_{\alpha\dot{\beta}}\bar{\chi}_{e}^{\dot{\beta}}\right]+\nonumber \\
\nonumber \\
 &  & +\dfrac{2i}{3}\epsilon^{abcd}E_{a}E_{b}E_{c}E^{\alpha}\left(\sigma_{d}\right)_{\alpha\dot{\alpha}}\left[\bar{\chi}^{\dot{\alpha}}F+i\bar{C}\left(\bar{\sigma}_{e}\right)^{\dot{\alpha}\beta}\chi_{\beta}^{e}\right]+\nonumber \\
\nonumber \\
 &  & +E_{a}E_{b}E_{c}E_{d}\epsilon^{abcd}\left[\dfrac{1}{2}C^{e}\phantom{}_{e}\bar{C}+\dfrac{1}{2}C\bar{C}^{e}\phantom{}_{e}+\dfrac{i}{2}\bar{\chi}_{\dot{\alpha}}^{e}\left(\bar{\sigma}_{e}\right)^{\dot{\alpha}\alpha}\chi_{\alpha}-\dfrac{i}{2}\bar{\chi}_{\dot{\alpha}}\left(\bar{\sigma}_{e}\right)^{\dot{\alpha}\alpha}\chi_{\alpha}^{e}+F\bar{F}\right].\label{eq:lagrangian}
\end{eqnarray}

An important comment is that, though the unfolded on-shell system
from Section \ref{sec:5} describes free dynamics of massless scalar
supermultiplet, its off-shell modification represents just an infinite
set of constraints. The form of these constraints is independent of
whether the model is free or nonlinear. As a result, it can be used
for description of a massive interacting theory. Nonlinear (starting
from cubic) representatives of $Q$-cohomology determine Lagrangians
with interactions. In particular, if $W=W\left(C\right)$ depends
only on $C$ (respectively $\bar{W}=\bar{W}\left(\bar{C}\right)$),
\eqref{eq:gen_lagr} describes the superpotential.

\subsection{Lagrangian as integral form\label{sub:integral_form}}

As explained in Section \ref{sec:3}, a superspace Lagrangian can
also be formulated as an integral form. This can be written as 
\begin{equation}
\mathcal{L}=E_{a_{1}}...E_{a_{m}}\delta^{2}\left(E_{\alpha}\right)\delta^{2}\left(\bar{E}_{\dot{\alpha}}\right)\ell^{[a_{1}...a_{m}]},\label{eq:cov_int_form}
\end{equation}
where $\ell^{[a_{1}...a_{m}]}$ is some Lorentz-covariant 0-form built
from the supermultiplet fields.

Applying the homotopy lemma, in spinor notations we have 
\begin{equation}
\mathcal{H}\left(E_{\alpha_{1}\dot{\alpha}_{1}}...E_{\alpha_{m}\dot{\alpha}_{m}}\delta^{2}\left(E_{\beta}\right)\delta^{2}\left(\bar{E}_{\dot{\beta}}\right)\ell^{\alpha_{1}...\alpha_{m},\dot{\alpha}_{1}...\dot{\alpha}_{m}}\right)=4\mathcal{L}+m\mathcal{L}-m\mathcal{L}-m\mathcal{L}=0,
\end{equation}
which implies $m=4$. So the only nonzero cohomology of $Q_{3}$ is
\begin{equation}
\mathcal{L}=E^{\alpha}\phantom{}_{\dot{\alpha}}E^{\beta\dot{\alpha}}E_{\beta\dot{\beta}}E_{\alpha}\phantom{}^{\dot{\beta}}\delta^{2}\left(E_{\gamma}\right)\delta^{2}\left(\bar{E}_{\dot{\gamma}}\right)\ell.\label{eq:L_int_form}
\end{equation}
It is elementary to see that this Lagrangian is $Q$-closed.

To analyze whether or not such actions contain trivial parts one has
to consider $Q$-images of the expressions containing derivatives
of delta-functions. Indeed, for 
\begin{equation}
\mathcal{F}=E^{\alpha}\phantom{}_{\dot{\alpha}}E^{\beta\dot{\alpha}}E_{\beta\dot{\beta}}E_{\alpha}\phantom{}^{\dot{\beta}}\delta_{\delta}'\left(E_{\gamma}\right)\delta^{2}\left(\bar{E}_{\dot{\gamma}}\right)f^{\delta}+h.c.
\end{equation}
we obtain 
\begin{equation}
Q\mathcal{F}=\left(Q_{2}^{+}+Q_{2}^{-}\right)\mathcal{F}=\sqrt{2}E^{\alpha}\phantom{}_{\dot{\alpha}}E^{\beta\dot{\alpha}}E_{\beta\dot{\beta}}E_{\alpha}\phantom{}^{\dot{\beta}}\delta^{2}\left(E_{\gamma}\right)\delta^{2}\left(\bar{E}_{\dot{\gamma}}\right)\left(\hat{q}^{\delta}f_{\delta}+h.c.\right)\,.
\end{equation}
As a result, \eqref{eq:L_int_form} describes trivial Lagrangians
if $\ell=\hat{q}^{\alpha}f_{\alpha}+h.c.$ for some $f_{\alpha}$.
In particular, Lagrangians with $\ell=\hat{q}^{\alpha\dot{\alpha}}f_{\alpha\dot{\alpha}}$
are trivial because, as follows from \eqref{eq:{q,q}_spin}, in this
case 
\begin{equation}
\ell=i\hat{q}^{\alpha}\left(\hat{\bar{q}}^{\dot{\alpha}}f_{\alpha\dot{\alpha}}\right)+i\hat{\bar{q}}^{\dot{\alpha}}\left(\hat{q}^{\alpha}f_{\alpha\dot{\alpha}}\right).
\end{equation}
In tensor indices the Lagrangian reads as 
\begin{equation}
\mathcal{L}=\epsilon^{abcd}E_{a}E_{b}E_{c}E_{d}\delta^{2}\left(E_{\alpha}\right)\delta^{2}\left(\bar{E}_{\dot{\alpha}}\right)\ell.\label{eq:L_int_tens}
\end{equation}
The Lagrangian which reproduces the free Salam-Strathdee Lagrangian
\cite{18 Wess Bagger Supersymmetry and Supergravity} is 
\begin{equation}
\mathcal{L^{WB}}=\epsilon^{abcd}E_{a}E_{b}E_{c}E_{d}\delta^{2}\left(E_{\alpha}\right)\delta^{2}\left(\bar{E}_{\dot{\alpha}}\right)\left(\bar{C}C\right),\label{eq:lagr_W-B}
\end{equation}
as follows from the formula \eqref{eq:DC=00003D0_DF=00003D0} which
implies that $C\left(z\right)$ is a chiral superfield.

To relate Lagrangians \eqref{eq:L_int_tens} and \eqref{eq:gen_lagr}
one can choose the integration surface for \eqref{eq:gen_lagr} as
\begin{equation}
x^{\underline{m}}=f^{\underline{m}}(t^{\underline{n}}),\qquad\theta^{\underline{\mu}}=\varphi^{\underline{\mu}}\left(t^{\underline{n}}\right),\qquad\bar{\theta}^{\dot{\underline{\mu}}}=\bar{\varphi}^{\dot{\underline{\mu}}}\left(t^{\underline{n}}\right)\,,\label{eq:surface_formula}
\end{equation}
where coordinates $t^{\underline{n}}$, $\underline{n}=1,...4$ are
even, $f^{\underline{m}}(t^{\underline{n}})$ are even and $\varphi^{\underline{\mu}}\left(t^{\underline{n}}\right),\,\bar{\varphi}^{\dot{\underline{\mu}}}\left(t^{\underline{n}}\right)$
are odd. Extending $t^{\underline{n}}$ by odd variables $\lambda^{\underline{\mu}},\bar{\lambda}^{\dot{\underline{\mu}}}$,
$\underline{\mu},\dot{\underline{\mu}}=1,2$ so that $\left\{ t^{\underline{n}},\lambda^{\underline{\mu}},\bar{\lambda}^{\dot{\underline{\mu}}}\right\} $
provide a full set of superspace coordinates, we extend \eqref{eq:surface_formula}
to 
\begin{equation}
x^{\underline{m}}=f^{\underline{m}}+i\varphi^{\underline{\mu}}\left(\sigma^{\underline{m}}\right)_{\underline{\mu}\dot{\underline{\mu}}}\bar{\lambda}^{\dot{\underline{\mu}}}-i\lambda^{\underline{\mu}}\left(\sigma^{\underline{m}}\right)_{\underline{\mu}\dot{\underline{\mu}}}\bar{\varphi}^{\dot{\underline{\mu}}},\qquad\theta^{\underline{\mu}}=\varphi^{\underline{\mu}}+\lambda^{\underline{\mu}},\qquad\bar{\theta}^{\dot{\underline{\mu}}}=\bar{\varphi}^{\dot{\underline{\mu}}}+\bar{\lambda}^{\dot{\underline{\mu}}}\,.\label{eq:zamen_perem}
\end{equation}
Substitution of \eqref{eq:zamen_perem} into $S=\int\mathcal{L}$
\eqref{eq:L_int_tens} and integration over $\lambda^{\underline{\mu}},\bar{\lambda}^{\dot{\underline{\mu}}}$
represents the action as an integral over the even surface \eqref{eq:surface_formula}
of the Lagrangian \eqref{eq:gen_lagr}, where $W=\hat{\bar{q}}_{\dot{\alpha}}\hat{\bar{q}}^{\dot{\alpha}}\ell$
and $\bar{W}=\hat{q}^{\alpha}\hat{q}_{\alpha}\ell$ (plus $Q$-exact
terms). In the process, 1-forms $E_{a},$ $E_{\alpha},$ $\bar{E}_{\dot{\alpha}}$
from \eqref{eq:L_int_tens} get transformed into Cartesian 1-forms
\eqref{eq:cartes_coord} in the resulting Lagrangian \eqref{eq:gen_lagr}
\begin{equation}
\tilde{E}^{a}=df^{a}(t)+d\varphi^{\underline{\mu}}\left(t\right)\left(i\bar{\varphi}^{\dot{\underline{\mu}}}\left(t\right)\left(\sigma^{a}\right)_{\underline{\mu}\dot{\underline{\mu}}}\right)+d\bar{\varphi}_{\dot{\underline{\mu}}}(t)\left(i\varphi^{\underline{\mu}}\left(t\right)\left(\sigma^{a}\right)_{\underline{\mu}\dot{\underline{\nu}}}\epsilon^{\dot{\underline{\mu}}\dot{\underline{\nu}}}\right),\quad\tilde{E}^{\alpha}=d\varphi^{\alpha}\left(t\right),\quad\tilde{\bar{E}}_{\dot{\alpha}}=d\bar{\varphi}{}_{\dot{\alpha}}\left(t\right).\label{eq:E_cartes}
\end{equation}
For the free Salam-Strathdee Lagrangian \eqref{eq:lagr_W-B}, this
gives the unfolded Wess-Zumino action with Lagrangian \eqref{eq:lagrangian}
hence showing their equivalence.

However, superpotentials are not represented in the integral form
\eqref{eq:L_int_tens}. This is expected since they represent chiral
functions to be integrated over chiral subspaces \cite{18 Wess Bagger Supersymmetry and Supergravity}.
In our approach such terms also are most conveniently represented
in the form intermediate between the 4-form Lagrangians and integral-form
Lagrangians, i.e. as integrals over chiral superspace.

\subsection{Chiral superspace}

To introduce superpotentials we introduce the following chiral integral
forms, 
\begin{equation}
\Lambda=\delta^{2}\left(E_{\alpha}\right)E_{a_{1}}...E_{a_{m}}\bar{E}_{\dot{\alpha}}...\bar{E}_{\dot{\alpha}}W^{[a_{1}...a_{m}]\dot{\alpha}(n)},\label{eq:chir_form}
\end{equation}
where $W$ is the Lorentz-covariant 0-form built from chiral functions\textbf{
$C^{a(k)}$} and $\bar{F}^{a(k)}$ (so, $Q_{2}^{-}\Lambda=0$). Such
forms are integrable over chiral superspace $\mathbb{C}^{m+n|2}$
in a standard way described in Section \ref{sec:3}. Here $\bar{E}_{\dot{\alpha}}$
without $\delta$-functions describe the pullback of the respective
1-forms to the chiral superspace.

Let us explore equation $\mathcal{H}\Lambda_{i}=0$ in spinor notations,
where $\Lambda_{i}$ from \eqref{eq:chir_form} contains $i$ factors
of $E^{\alpha\dot{\alpha}}$. For 
\begin{eqnarray}
 &  & \Lambda_{0}=\delta^{2}\left(E_{\alpha}\right)\bar{E}^{\dot{\alpha}}...\bar{E}^{\dot{\alpha}}W_{\dot{\alpha}(m)},\label{eq:HW0}\\
\nonumber \\
 &  & \Lambda_{1}=\delta^{2}\left(E_{\alpha}\right)\bar{E}^{\dot{\alpha}}...\bar{E}^{\dot{\alpha}}E^{\beta\dot{\beta}}W_{\beta,\dot{\beta},\dot{\alpha}(m)},\\
\nonumber \\
 &  & \Lambda_{2}=\delta^{2}\left(E_{\alpha}\right)\bar{E}^{\dot{\alpha}}...\bar{E}^{\dot{\alpha}}\left(E^{\beta}\phantom{}_{\dot{\beta}}E^{\gamma\dot{\beta}}W_{\beta\gamma,\dot{\alpha}(m)}+E^{\beta\dot{\beta}}E_{\beta}\phantom{}^{\dot{\gamma}}W_{\dot{\beta}\dot{\gamma},\dot{\alpha}(m)}\right),\\
\nonumber \\
 &  & \Lambda_{3}=\delta^{2}\left(E_{\alpha}\right)E^{\beta}\phantom{}_{\dot{\gamma}}E^{\gamma\dot{\gamma}}E_{\gamma}\phantom{}^{\dot{\beta}}\bar{E}^{\dot{\alpha}}...\bar{E}^{\dot{\alpha}}W_{\beta,\dot{\beta},\dot{\alpha}(m)},
\end{eqnarray}
the equation $\mathcal{H}\Lambda=0$ has nontrivial solutions (with
arbitrary 0-forms $W$) only at $m=0$ . This is easy to understand
by noting that, because $\delta^{2}\left(E_{\alpha}\right)\bar{E}^{\dot{\alpha}}=Q_{3}\left(\frac{i}{4}\delta'_{\beta}\left(E_{\alpha}\right)E^{\beta\dot{\alpha}}\right)$
all such $\Lambda$, being $Q_{3}$-closed due to $\delta$-functions,
are not $Q_{3}$-exact only if they are independent of $\bar{E}^{\dot{\alpha}}$.

However, for 
\begin{equation}
\Lambda_{4}=\delta^{2}\left(E_{\alpha}\right)E^{\beta}\phantom{}_{\dot{\beta}}E^{\gamma\dot{\beta}}E_{\gamma\dot{\gamma}}E_{\beta}\phantom{}^{\dot{\gamma}}\bar{E}^{\dot{\alpha}}...\bar{E}^{\dot{\alpha}}W_{\dot{\alpha}(m)}\,,\label{eq:HW4}
\end{equation}
$\mathcal{H}\Lambda_{4}=-2m\Lambda_{4}+4\Lambda_{4}-4\Lambda_{4}+2m\Lambda_{4}\equiv0\,.$
Hence, $\Lambda_{4}$ \eqref{eq:HW4} can represent $Q_{3}$-cohomology
at any $m$. Thus $Q_{3}$-cohomology in the class of chiral integral
forms is represented by 
\begin{eqnarray}
 &  & \Lambda_{0}=\delta^{2}\left(E_{\alpha}\right)W,\label{eq:W0}\\
\nonumber \\
 &  & \Lambda_{1}=\delta^{2}\left(E_{\alpha}\right)E^{\beta\dot{\beta}}W_{\beta\dot{\beta}},\\
\nonumber \\
 &  & \Lambda_{2}=\delta^{2}\left(E_{\alpha}\right)\left(E^{\beta}\phantom{}_{\dot{\beta}}E^{\gamma\dot{\beta}}W_{\beta\gamma}+E^{\beta\dot{\beta}}E_{\beta}\phantom{}^{\dot{\gamma}}\bar{W}_{\dot{\beta}\dot{\gamma}}\right),\\
\nonumber \\
 &  & \Lambda_{3}=\delta^{2}\left(E_{\alpha}\right)E^{\beta}\phantom{}_{\dot{\gamma}}E^{\gamma\dot{\gamma}}E_{\gamma}\phantom{}^{\dot{\beta}}W_{\beta\dot{\beta}},\label{eq:W3}\\
\nonumber \\
 &  & \Lambda_{4}=\delta^{2}\left(E_{\alpha}\right)E^{\beta}\phantom{}_{\dot{\beta}}E^{\gamma\dot{\beta}}E_{\gamma\dot{\gamma}}E_{\beta}\phantom{}^{\dot{\gamma}}\bar{E}^{\dot{\alpha}}...\bar{E}^{\dot{\alpha}}W_{\dot{\alpha}(m)}.\label{eq:W4}
\end{eqnarray}
To be integrated over $\mathbb{C}^{4|2}$, the only appropriate expression
from \eqref{eq:W0}-\eqref{eq:W4} is 
\begin{equation}
\mathcal{\mathcal{L}}_{0}=\delta^{2}\left(E_{\alpha}\right)E^{\beta}\phantom{}_{\dot{\beta}}E^{\gamma\dot{\beta}}E_{\gamma\dot{\gamma}}E_{\beta}\phantom{}^{\dot{\gamma}}W.\label{eq:L0_chir}
\end{equation}
Note that $G\left(\delta^{2}\left(E_{\alpha}\right)\right)=-4$, because
the $\delta$-function has the degree of homogeneity $-2$ and $G\left(E_{\alpha}\right)=2$,
as a result $G\left(\mathcal{\mathcal{L}}_{0}\right)=0$\textbf{.}
To single out trivial Lagrangians, we write down Eqs.~\eqref{eq:triv_L_1}-\eqref{eq:triv_L_2}
for 
\begin{equation}
\mathcal{F}_{-1}=\delta^{2}\left(E_{\alpha}\right)E^{\beta}\phantom{}_{\dot{\gamma}}E^{\gamma\dot{\gamma}}E_{\gamma}\phantom{}^{\dot{\beta}}f_{\beta\dot{\beta}},\quad G\left(\mathcal{F}_{-1}\right)=-1,\label{eq:F_-1}
\end{equation}
which represents the only expression from those in \eqref{eq:W0}-\eqref{eq:W4},
whose $Q$-image can contribute to \eqref{eq:L0_chir}. This gives
\begin{eqnarray}
 &  & Q_{3}\mathcal{F}_{-1}=0,\label{eq:Q3_F-1=00003D0}\\
\nonumber \\
 &  & \left(Q_{2}^{+}+Q_{2}^{-}\right)\mathcal{F}_{-1}+Q_{3}\mathcal{F}_{-2}=0,\label{eq:Q2_F-1=00003D0}\\
\nonumber \\
 &  & Q_{1}\mathcal{F}_{-1}+\left(Q_{2}^{+}+Q_{2}^{-}\right)\mathcal{F}_{-2}+Q_{3}\mathcal{F}_{-3}=\delta^{2}\left(E_{\alpha}\right)E^{\beta}\phantom{}_{\dot{\beta}}E^{\gamma\dot{\beta}}E_{\gamma\dot{\gamma}}E_{\beta}\phantom{}^{\dot{\gamma}}W.\label{eq:Q1_F-1=00003D0}
\end{eqnarray}
We immediately conclude that $\mathcal{F}_{-2}$ with $G=-2$ here
can only have the form 
\[
\mathcal{F}_{-2}=\delta'_{\delta}\left(E_{\alpha}\right)E^{\beta}\phantom{}_{\dot{\beta}}E^{\gamma\dot{\beta}}E_{\gamma\dot{\gamma}}E_{\beta}\phantom{}^{\dot{\gamma}}f^{\delta}.
\]
(\textbf{$G$$\left(\delta'_{\beta}\left(E_{\alpha}\right)\right)=-6$
}as follows from the definition \textbf{$E_{\gamma}\delta'_{\beta}\left(E_{\alpha}\right)=\epsilon_{\gamma\beta}\delta^{2}\left(E_{\alpha}\right)$
}and \textbf{$G\left(\delta\left(E_{\alpha}\right)\right)=-4$}.)
Eq.~\eqref{eq:Q3_F-1=00003D0} is satisfied as $\mathcal{F}_{-1}\in H\left(Q_{3}\right)$.
Term with $\mathcal{F}_{-1}$ in \eqref{eq:Q2_F-1=00003D0} is zero
because \eqref{eq:F_-1} contains $\delta$-function and $f_{\beta\dot{\beta}}$
is built from \textbf{$C^{a(k)}$} and $\bar{F}^{a(k)}$, so $\mathcal{F}_{-2}=0$.
Finally, Eq.~\eqref{eq:Q1_F-1=00003D0} has a solution 
\[
\mathcal{F}_{-3}=0,\quad W=\frac{1}{8}\hat{q}_{\alpha\dot{\alpha}}f^{\alpha\dot{\alpha}}\,.
\]
We conclude that \eqref{eq:L0_chir} describes nontrivial Lagrangians
only if $W\neq\hat{q}_{\alpha\dot{\alpha}}f^{\alpha\dot{\alpha}}$
for some $f^{\alpha\dot{\alpha}}$. Also, analogously to Subsection
\ref{sub:integral_form}, we have to consider expressions with derivatives
of $\delta$-function whose $Q$-images can lead to trivial Lagrangians
in \eqref{eq:L0_chir}. It is easy to see that the only appropriate
elements from $Ker\left(Q_{3}\right)$ are 
\begin{eqnarray*}
 &  & K_{1}=\delta'_{\beta}\left(E_{\alpha}\right)\bar{E}^{\dot{\alpha}}...\bar{E}^{\dot{\alpha}}\bar{k}^{\beta}\phantom{}_{\dot{\alpha}(4)},\\
\\
 &  & K_{2}=\delta'_{\beta}\left(E_{\alpha}\right)E^{\gamma\dot{\alpha}}E_{\gamma}\phantom{}^{\dot{\beta}}\bar{E}_{\dot{\alpha}}\bar{E}_{\dot{\beta}}\bar{k}^{\beta}.
\end{eqnarray*}
However the former has $G\left(K_{1}\right)=2$ while the latter has
$G\left(K_{1}\right)=0$. Since $G\left(\mathcal{\mathcal{L}}_{0}\right)=0$
for \eqref{eq:L0_chir}, the system \eqref{eq:triv_L_1}-\eqref{eq:triv_L_2}
admits no nonzero solutions in these cases.

Next, besides $Q_{3}\mathcal{\mathcal{L}}_{0}=0$, $\mathcal{L}_{0}$
\eqref{eq:L0_chir} obeys 
\begin{eqnarray}
 &  & Q_{2}^{-}\mathcal{\mathcal{L}}_{0}=0,\\
\nonumber \\
 &  & Q_{2}^{+}\mathcal{\mathcal{L}}_{0}=0,\\
\nonumber \\
 &  & Q_{1}\mathcal{\mathcal{L}}_{0}=0.
\end{eqnarray}
The first equation holds because $W$ in \eqref{eq:L0_chir} is built
from chiral \textbf{$C^{a(k)}$} and $\bar{F}^{a(k)}$. The second
one holds due to the $\delta$-function. The third one is true because
$\mathcal{\mathcal{L}}_{0}$ contains the maximal number of $E^{\alpha\dot{\alpha}}$.
So $\mathcal{L}_{0}$ \eqref{eq:L0_chir} is $Q$-closed and hence
represents the general form of an unfolded chiral Lagrangian.\textbf{
}(Note that \eqref{eq:W0}-\eqref{eq:W3} with nonzero $W$ are not
$Q$-closed, because they contain less than four $E^{\alpha\dot{\alpha}}$
and hence are not annihilated by $Q_{1}=\tfrac{1}{2}E^{\alpha\dot{\alpha}}\hat{q}_{\alpha\dot{\alpha}}$.)

In tensor notations, general chiral Lagrangian is 
\begin{equation}
\mathcal{\mathcal{L}}=\delta^{2}\left(E_{\alpha}\right)\epsilon^{abcd}E_{a}E_{b}E_{c}E_{d}W\,,\label{eq:gen_chiral_L}
\end{equation}
where Lorentz-invariant 0-form $W$ is built from \textbf{$C^{a(k)}$}
and $\bar{F}^{a(k)}$, and $W\neq\hat{q}^{a}f_{a}$. For the Lagrangian
to be real, \eqref{eq:gen_chiral_L} should be supplemented by the
complex conjugated expression to be integrated over antichiral superspace
with the 0-form $\bar{W}$ built from \textbf{$\bar{C}^{a(k)}$} and
$F^{a(k)}$. Combination of \eqref{eq:gen_chiral_L} (plus conjugated
expression) and \eqref{eq:L_int_tens} gives the unfolded superspace
action of the interacting theory.

To reproduce superpotential of the Wess-Zumino model \cite{18 Wess Bagger Supersymmetry and Supergravity}
we choose $W=kC+\dfrac{m}{2}CC+\dfrac{g}{3}CCC$ with arbitrary constants
$k$, $m$, $g$. Then superpotential takes the form 
\begin{equation}
\mathcal{\mathcal{L}}=\int\delta^{2}\left(E_{\alpha}\right)\epsilon^{abcd}E_{a}E_{b}E_{c}E_{d}\left(kC+\dfrac{m}{2}CC+\dfrac{g}{3}CCC\right)+h.c.\label{eq:L_pot}
\end{equation}

To write a full action containing both kinetic term and superpotential
of the Wess-Zumino model in the chiral form we set $W=-\dfrac{1}{16}C\bar{F}+kC+\dfrac{m}{2}CC+\dfrac{g}{3}CCC$.
This gives 
\begin{equation}
S=\int\delta^{2}\left(E_{\alpha}\right)\epsilon^{abcd}E_{a}E_{b}E_{c}E_{d}\left(-\dfrac{1}{16}C\bar{F}+kC+\dfrac{m}{2}CC+\dfrac{g}{3}CCC\right)+h.c.\label{eq:chiral_WZ}
\end{equation}
The kinetic term $-\dfrac{1}{16}C\bar{F}$ in \eqref{eq:chiral_WZ}
results from integration of \eqref{eq:lagr_W-B} over $\bar{\theta}_{\dot{\alpha}}$
taking into account that from \eqref{eq:*off-shell_1}-\eqref{eq:*off-shell_2}
it follows that $\bar{F}=2\bar{D}\bar{D}\bar{C}$.

As in Section \ref{sub:integral_form}, one can map solution \eqref{eq:gen_chiral_L}
to the 4-superform \eqref{eq:gen_lagr}. The only difference is that
instead of \eqref{eq:zamen_perem} even surfaces of the chiral superspace
$\left(x^{m},\theta^{\mu}\right)\rightarrow\left(t^{m},\lambda^{\mu}\right)$
are parametrized as 
\begin{equation}
x^{m}=f^{m}\left(t\right)+i\left(\varphi^{\mu}\left(t\right)+\lambda^{\mu}\right)\left(\sigma^{m}\right)_{\mu\dot{\mu}}\bar{\varphi}^{\dot{\mu}}\left(t\right),\qquad\theta^{\mu}=\varphi^{\mu}\left(t\right)+\lambda^{\mu}\,.
\end{equation}

Integration over $\lambda^{\mu}$, gives Lagrangian \eqref{eq:gen_lagr}
with the same function $W$ as in \eqref{eq:gen_chiral_L}. Analogously
to Section \ref{sub:integral_form}, the resulting Lagrangian is integrated
over the surface \eqref{eq:surface_formula} with 1-forms \eqref{eq:E_cartes}.

Expressions (\ref{eq:gen_lagr}), (\ref{eq:L_int_tens}) and (\ref{eq:gen_chiral_L})
give the most general unfolded Lagrangians which can be written for
the Wess-Zumino model as a 4-superform, integral form or a chiral
integral form respectively. Besides the standard Wess-Zumino and Salam-Strathdee
Lagrangians, they also describe the higher-derivative Lagrangians.
Namely, by virtue of Eqs.~(\ref{eq: off-shell_1})-(\ref{eq:off-shell_2}),
the rank $k$ tensors $C^{a(k)}$, $\chi_{\alpha}^{a(k)}$ and $F^{a(k)}$
describe the $k$-th derivatives of the dynamical fields $C$, $\chi_{\alpha}$
and $F$. Plugging the higher-rank tensors into (\ref{eq:gen_lagr}),
(\ref{eq:L_int_tens}) or (\ref{eq:gen_chiral_L}) gives the higher-derivative
unfolded actions. Unfolded Lagrangians contain all possible ordinary
Lagrangians that can be written for the $4d$ Wess-Zumino model. To
obtain a conventional field-theoretic Lagrangian from the unfolded,
one has to express all auxiliary fields $C^{a(k)},$ $\bar{C}^{a(k)},$
$\chi_{\alpha}^{a(k)},$ $\bar{\chi}_{\dot{\alpha}}^{a(k)},$ $F^{a(k)},$
$\bar{F}^{a(k)}$ with $k\geqslant1$ in terms of the derivatives
of the dynamical fields using unfolded equations (\ref{eq: off-shell_1})-(\ref{eq:off-shell_2}),
(\ref{eq:*off-shell_1})-(\ref{eq:*off-shell_2}). Then, fixing an
integration surface, the substitution of the resulting expressions
for instance into (\ref{eq:gen_lagr}) gives an ordinary space-time
action.

In particular, doing this for (\ref{eq:lagrangian}) and choosing
Minkowski space as an integration surface we see that the integral
of (\ref{eq:lagrangian}) reproduces the component action of the free
chiral supermultiplet \cite{18 Wess Bagger Supersymmetry and Supergravity}.
Alternatively, one can use Lagrangians (\ref{eq:lagr_W-B}) or (\ref{eq:chiral_WZ}),
arriving at the standard Salam-Strathdee superfield action. Generally,
being manifestly supersymmetric the unfolded superform action leads
directly to the component action.

\section{Conclusion}

In this paper, unfolded off-shell formulation of the free massless
scalar supermultiplet is presented and the system of equations, that
determines all superfield Lagrangians of the model, is derived and
analyzed. The particular solutions leading to superfield actions of
the Wess-Zumino model in the form of integrals of a 4-superform, integral
form and chiral integral form are obtained. Explicit relations between
these forms of superspace actions are established. It is shown in
particular how usual superspace action for the Wess-Zumino model can
be rewritten as an integral of a 4-superform.

In some sense, the construction of a chiral action is intermediate
between the one of Section \ref{sub:4-form} and that of Section \ref{sub:integral_form}.
In fact, this is a particular example of a very general phenomenon
that the full action may have a form of integral over (super)manifolds
of different dimensions. As long as even dimension is kept fixed,
integration over supercoordinates will result in one or another space-time
action. We expect that in more complicated theories like higher-spin
theories and their further multiparticle extensions, invariant functionals
resulting from integrals over space-times with different even dimensions
may all contribute to the final result.

Being based on unfolded dynamics, the proposed method is most general,
providing maximal flexibility in the construction of supersymmetric
actions. Applied to on-shell unfolded system it provides a systematic
tool for the analysis of on-shell counterterms in supersymmetric systems,
the issue which was extensively studied during the recent years \cite{Berkovits,Bossard,Bossard2,Movshev,Chang 1,Chang 2}.
It would be interesting to explore its applications to more complicated
models with extended SUSY and, in the first place, to the theories
whose manifestly supersymmetric formulation is yet lacking, like $\mathcal{N}=1$,
$D=10$ or $\mathcal{N}=4$, $D=4$ super Yang-Mills theories. It
would also be interesting to clarify the relation of our approach
to the harmonic superspace approach \cite{Galperin:1984av}.

\section*{Acknowledgments}

The authors are grateful to A. Barvinsky, V. Didenko and B. Voronov
for useful comments. This research was supported in part by RFBR Grants
No. 11-02-00814-a, 12-02-31838. N.M. acknowledges financial support
from Dynasty Foundation.

\section*{Appendix A\label{A}}

We work with 4-dimensional Minkowski space with coordinates $x^{\underline{m}},\:\underline{m}=0...3$
and the superspace $\mathbb{R}^{4|4}$ with coordinates $z^{\underline{M}}=\left(x^{\underline{m}},\theta^{\underline{\mu}},\bar{\theta}^{\dot{\underline{\mu}}}\right);\:\underline{m}=0...3;\:\underline{\mu,}\underline{\dot{\,\mu}}=1,2$,
which are denoted by the underlined letters from the middle of Latin
and Greek alphabets. Supervielbeins $E_{a},E_{\alpha},\bar{E}_{\dot{\alpha}}$
relate the base indices to the indices of the flat fiber space, denoted
by the letters from the beginning of the respective alphabets: $\left(x^{a},\theta^{\alpha},\bar{\theta}^{\dot{\alpha}}\right);\: a=0...3;\:\alpha,\dot{\alpha}=1,2$.

The fiber space Minkowski metric is $\eta_{ab}=diag\left\{ 1,-1,-1,-1\right\} $.
We use condensed notations for symmetrized fiber indices writing $a(k)$
instead of $\left(a_{1}...a_{k}\right)$. Indices in brackets $\left[a_{1}...a_{k}\right]$
are antisymmetrized. Spinorial indices are raised and lowered by the
matrices

\[
\epsilon^{\alpha\beta}=\epsilon^{\dot{\alpha}\dot{\beta}}=\left\Vert \begin{array}{cc}
0 & 1\\
-1 & 0
\end{array}\right\Vert ,\qquad\epsilon_{\alpha\beta}=\epsilon_{\dot{\alpha}\dot{\beta}}=\left\Vert \begin{array}{cc}
0 & -1\\
1 & 0
\end{array}\right\Vert ,\eqno(A.1)
\]

\[
\xi_{\alpha}=\epsilon_{\alpha\beta}\xi^{\beta},\qquad\xi^{\alpha}=\epsilon^{\alpha\beta}\xi_{\beta},\qquad\bar{\xi}^{\dot{\alpha}}=\epsilon^{\dot{\alpha}\dot{\beta}}\bar{\xi}_{\dot{\beta}},\qquad\bar{\xi}_{\dot{\alpha}}=\epsilon_{\dot{\alpha}\dot{\beta}}\bar{\xi}^{\dot{\beta}}.\eqno(A.2)
\]
$\sigma$-matrices are 
\[
\left(\sigma^{0}\right)_{\alpha\dot{\beta}}=\left\Vert \begin{array}{cc}
1 & 0\\
0 & 1
\end{array}\right\Vert ,\;\left(\sigma^{1}\right)_{\alpha\dot{\beta}}=\left\Vert \begin{array}{cc}
0 & 1\\
1 & 0
\end{array}\right\Vert ,\;\left(\sigma^{2}\right)_{\alpha\dot{\beta}}=\left\Vert \begin{array}{cc}
0 & -i\\
i & 0
\end{array}\right\Vert ,\;\left(\sigma^{3}\right)_{\alpha\dot{\beta}}=\left\Vert \begin{array}{cc}
1 & 0\\
0 & -1
\end{array}\right\Vert .\eqno(A.3)
\]
Also we use anti-Hermitian matrices 
\[
\left(\sigma_{ab}\right)_{\alpha}{}^{\beta}=\frac{1}{2}\left(\left(\sigma_{a}\right)_{\alpha\dot{\alpha}}\left(\bar{\sigma}_{b}\right)^{\dot{\alpha}\beta}-\left(\sigma_{b}\right)_{\alpha\dot{\alpha}}\left(\bar{\sigma}_{a}\right)^{\dot{\alpha}\beta}\right),\;\left(\bar{\sigma}_{ab}\right)^{\dot{\alpha}}{}_{\dot{\beta}}=\frac{1}{2}\left(\left(\bar{\sigma}_{a}\right)^{\dot{\alpha}\alpha}\left(\bar{\sigma}_{b}\right)_{\alpha\dot{\beta}}-\left(\bar{\sigma}_{a}\right)^{\dot{\alpha}\alpha}\left(\bar{\sigma}_{b}\right)_{\alpha\dot{\beta}}\right).\eqno(A.4)
\]
Their tensorial indices are lowered/raised by Levi-Civita symbol

\[
\left(\sigma^{ab}\right)_{\alpha}{}^{\beta}=\frac{i}{2}\epsilon^{abcd}\left(\sigma_{cd}\right)_{\alpha}{}^{\beta},\qquad\left(\bar{\sigma}^{ab}\right)^{\dot{\alpha}}{}_{\dot{\beta}}=-\frac{i}{2}\epsilon^{abcd}\left(\bar{\sigma}_{cd}\right)^{\dot{\alpha}}{}_{\dot{\beta}}.\eqno(A.5)
\]
Following well-known relations are used

\[
T_{\alpha\beta\gamma...}-T_{\beta\alpha\gamma...}=\epsilon_{\alpha\beta}T^{\delta}\phantom{}_{\delta\gamma...},\eqno(A.6)
\]

\[
\left(\sigma^{a}\right)_{\alpha\dot{\alpha}}\left(\bar{\sigma}_{a}\right)^{\beta\dot{\beta}}=2\delta_{\alpha}\phantom{}^{\beta}\delta_{\dot{\alpha}}\phantom{}^{\dot{\beta}},\eqno(A.7)
\]

\[
\left(\sigma_{a}\bar{\sigma}_{b}\right)_{\alpha}\phantom{}^{\beta}=\eta_{ab}\delta_{\alpha}\phantom{}^{\beta}+\left(\sigma_{ab}\right)_{\alpha}\phantom{}^{\beta}.\eqno(A.8)
\]
Using them one obtaines

\[
\left(\sigma_{a}\right)_{\alpha\dot{\alpha}}\left(\sigma_{b}\right)_{\beta\dot{\beta}}-\left(\sigma_{b}\right)_{\alpha\dot{\alpha}}\left(\sigma_{a}\right)_{\beta\dot{\beta}}=\left(\sigma_{ab}\right)_{\alpha\beta}\epsilon_{\dot{\alpha}\dot{\beta}}+\left(\bar{\sigma}_{ab}\right)_{\dot{\alpha}\dot{\beta}}\epsilon_{\alpha\beta},\eqno(A.9)
\]

\[
\left(\sigma_{a}\right)_{\alpha\dot{\alpha}}\left(\sigma_{b}\right)_{\beta\dot{\beta}}+\left(\sigma_{b}\right)_{\alpha\dot{\alpha}}\left(\sigma_{a}\right)_{\beta\dot{\beta}}=\eta_{ab}\epsilon_{\alpha\beta}\epsilon_{\dot{\alpha}\dot{\beta}}+\left(\sigma_{ca}\right)_{\alpha\beta}\left(\bar{\sigma}_{cb}\right)_{\dot{\alpha}\dot{\beta}},\eqno(A.10)
\]

\[
\left(\sigma^{a}\bar{\sigma}^{b}\sigma^{c}\right)_{\alpha\dot{\alpha}}-\left(\sigma^{c}\bar{\sigma}^{b}\sigma^{a}\right)_{\alpha\dot{\alpha}}=2i\epsilon^{abcd}\left(\sigma_{d}\right)_{\alpha\dot{\alpha}}.\eqno(A.11)
\]

\section*{Appendix B\label{B}}

\noindent Expanding $\hat{Q}\mathcal{L}=0$ for \eqref{eq:ansatz_L}
and \eqref{eq:Q_ost} into parts with different $G$-grade, one obtains
a following chain of equations

\begin{eqnarray*}
G=9: &  & Q_{3}\left[E_{a}E_{b}\left(\bar{\sigma}^{ab}\right)^{\dot{\alpha}\dot{\beta}}\bar{E}_{\dot{\alpha}}\bar{E}_{\dot{\beta}}\ell_{6}\right]=0,\\
 &  & Q_{3}\left[E_{a}E_{b}\left(\sigma^{ab}\right)^{\alpha\beta}E_{\alpha}E_{\beta}\bar{\ell}_{6}\right]=0,\\
\\
G=8: &  & Q_{2}^{-}\left[E_{a}E_{b}\left(\bar{\sigma}^{ab}\right)^{\dot{\alpha}\dot{\beta}}\bar{E}_{\dot{\alpha}}\bar{E}_{\dot{\beta}}\ell_{6}\right]=0,\\
 &  & Q_{2}^{+}\left[E_{a}E_{b}\left(\sigma^{ab}\right)^{\alpha\beta}E_{\alpha}E_{\beta}\bar{\ell}_{6}\right]=0,\\
 &  & Q_{3}\left[\epsilon^{abcd}E_{a}E_{b}E_{c}\bar{E}_{\dot{\alpha}}\left(\bar{\sigma}_{d}\right)^{\dot{\alpha}\alpha}\ell_{5}\phantom{}_{\alpha}\right]+Q_{2}^{+}\left[E_{a}E_{b}\left(\bar{\sigma}^{ab}\right)^{\dot{\alpha}\dot{\beta}}\bar{E}_{\dot{\alpha}}\bar{E}_{\dot{\beta}}\ell_{6}\right]=0,\\
 &  & Q_{3}\left[\epsilon^{abcd}E_{a}E_{b}E_{c}E^{\alpha}\left(\sigma_{d}\right)_{\alpha\dot{\alpha}}\bar{\ell}_{5}^{\dot{\alpha}}\right]+Q_{2}^{-}\left[E_{a}E_{b}\left(\sigma^{ab}\right)^{\alpha\beta}E_{\alpha}E_{\beta}\bar{\ell}_{6}\right]=0.\\
\\
G=7: &  & Q_{2}^{-}\left[\epsilon^{abcd}E_{a}E_{b}E_{c}\bar{E}_{\dot{\alpha}}\left(\bar{\sigma}_{d}\right)^{\dot{\alpha}\alpha}\ell_{5}\phantom{}_{\alpha}\right]+Q_{1}\left[E_{a}E_{b}\left(\bar{\sigma}^{ab}\right)^{\dot{\alpha}\dot{\beta}}\bar{E}_{\dot{\alpha}}\bar{E}_{\dot{\beta}}\ell_{6}\right]=0,\\
 &  & Q_{2}^{+}\left[\epsilon^{abcd}E_{a}E_{b}E_{c}E^{\alpha}\left(\sigma_{d}\right)_{\alpha\dot{\alpha}}\bar{\ell}_{5}^{\dot{\alpha}}\right]+Q_{1}\left[E_{a}E_{b}\left(\sigma^{ab}\right)^{\alpha\beta}E_{\alpha}E_{\beta}\bar{\ell}_{6}\right]=0,\\
 &  & Q_{3}\left[E_{a}E_{b}E_{c}E_{d}\epsilon^{abcd}\ell_{4}\right]+Q_{2}^{-}\left[\epsilon^{abcd}E_{a}E_{b}E_{c}E^{\alpha}\left(\sigma_{d}\right)_{\alpha\dot{\alpha}}\bar{\ell}_{5}^{\dot{\alpha}}\right]+\\
 &  & \quad+Q_{2}^{+}\left[\epsilon^{abcd}E_{a}E_{b}E_{c}\bar{E}_{\dot{\alpha}}\left(\bar{\sigma}_{d}\right)^{\dot{\alpha}\alpha}\ell_{5}\phantom{}_{\alpha}\right]=0.\\
\\
G=6: &  & Q_{2}^{-}\left[E_{a}E_{b}E_{c}E_{d}\epsilon^{abcd}\ell_{4}\right]+Q_{1}\left[\epsilon^{abcd}E_{a}E_{b}E_{c}\bar{E}_{\dot{\alpha}}\left(\bar{\sigma}_{d}\right)^{\dot{\alpha}\alpha}\ell_{5}\phantom{}_{\alpha}\right]=0,\\
 &  & Q_{2}^{+}\left[E_{a}E_{b}E_{c}E_{d}\epsilon^{abcd}\ell_{4}\right]+Q_{1}\left[\epsilon^{abcd}E_{a}E_{b}E_{c}E^{\alpha}\left(\sigma_{d}\right)_{\alpha\dot{\alpha}}\bar{\ell}_{5}^{\dot{\alpha}}\right]=0.\\
\\
G=5: &  & Q_{1}\left[E_{a}E_{b}E_{c}E_{d}\epsilon^{abcd}\ell_{4}\right]=0.
\end{eqnarray*}

\end{document}